\documentclass[10pt,conference]{IEEEtran}

\IEEEoverridecommandlockouts
\usepackage{cite}
\usepackage{amsmath,amssymb,amsfonts}
\usepackage{algorithmic}
\usepackage{graphicx}
\usepackage{textcomp}
\usepackage{xcolor}
\usepackage{colortbl}
\def\BibTeX{{\rm B\kern-.05em{\sc i\kern-.025em b}\kern-.08em
    T\kern-.1667em\lower.7ex\hbox{E}\kern-.125emX}}

\usepackage{subcaption}
\usepackage{tabularx}
\usepackage{booktabs} 
\usepackage{tabularx}
\usepackage{xspace}
\usepackage{tcolorbox}
\usepackage{dashrule}
\usepackage{enumitem}
\usepackage{fancybox}
\usepackage{setspace}
\usepackage{url}
\usepackage[normalem]{ulem}

\newcommand{\codetask}{code generation tasks\xspace}
\newcommand{\framework}{CodeVisionary\xspace}
\newcommand{\ie}{\textit{i}.\textit{e}.\xspace}

\newcommand{\moduleA}{Requirement-guided multi-dimensional context distillation stage\xspace}
\newcommand{\moduleB}{Fine-grained scoring and summarization stage\xspace}
\newcommand{\modulea}{requirement-guided multi-dimensional context distillation stage\xspace}
\newcommand{\moduleb}{fine-grained scoring and summarization stage\xspace}

\usepackage{multirow}

\begin{document}
\title{An Agent-based Evaluation Framework for Complex Code Generation}

\author{Anonymous Author(s)}

\author{
    Xinchen Wang\textsuperscript{1$\dagger$}, 
    Ruida Hu\textsuperscript{1$\dagger$}, 
    Pengfei Gao\textsuperscript{2}, 
    Chao Peng\textsuperscript{2}, 
    Cuiyun Gao\textsuperscript{1*} \\

    \thanks{$^{\dagger}$Work done during an internship at ByteDance.}
    \thanks{$^{*}$Corresponding author.}
    
    \textsuperscript{1}Harbin Institute of Technology, Shenzhen, China \\
    \textsuperscript{2}ByteDance, Beijing, China \\
    200111115@stu.hit.edu.cn, 200111107@stu.hit.edu.cn, \\ gaopengfei.se@bytedance.com, pengchao.x@bytedance.com, 
     gaocuiyun@hit.edu.cn
}

\maketitle

\begin{abstract}
\label{sec:abstract}
Large language models (LLMs) have demonstrated strong capabilities in code generation, underscoring the critical need for rigorous and comprehensive evaluation. 
Existing evaluation approaches fall into three categories, including human-centered, metric-based, and LLM-based. Considering that human-centered approaches are labour-intensive and metric-based ones overly rely on reference answers, LLM-based approaches are gaining increasing attention due to their stronger contextual understanding capabilities.
However, they generally evaluate the generated code based on static prompts, and tend to fail for complex code scenarios which typically involve multiple requirements and require more contextual information. In addition, these approaches lack fine-grained evaluation for complex code, resulting in limited explainability.

To mitigate the limitations, we propose \textbf{\framework}, the first agent-based evaluation framework for complex code generation.
\framework consists of two stages: \textbf{(1) \textit{\moduleA}}, which first formulates a detailed evaluation plan by decomposing task requirements, and then stepwise collects multi-dimensional contextual information for each requirement.
\textbf{(2) \textit{\moduleB}}, which defines self-directed and negotiation-based actions, allowing multiple judges to comprehend complex code from fine-grained and diverse viewpoints, and reach a consensus through discussion. A comprehensive evaluation report is also generated for enhanced explainability. For validation, we construct a new benchmark consisting of 363 samples spanning 37 coding scenarios and 23 programming languages.
Extensive experiments demonstrate that \framework achieves the best performance among three baselines for evaluating complex code generation, outperforming the best baseline with average improvements of 0.217, 0.163, and 0.141 in Pearson, Spearman, and Kendall-Tau coefficients, respectively.
The resources of \framework are available at \url{https://github.com/Eshe0922/CodeVisionary}.

\begin{IEEEkeywords}
Code generation evaluation, large language models, AI agent 
\end{IEEEkeywords}







\end{abstract}

\section{Introduction}
\label{sec:introduction}
With the rapid development of large language models (LLMs), these models have demonstrated promising results in code generation~\cite{codetask1, codetask2}. LLM-based code assistants, such as GitHub Copilot~\cite{copilot} and Cursor\cite{cursor}, have attracted a great number of users, effectively addressing their programming needs. Effectively evaluating the capabilities of LLMs in accomplishing \codetask is beneficial for identifying their shortcomings, ultimately enhancing the practical utility in real-world software development scenarios.
Current approaches for evaluating code generation can be categorized into human-centered, metric-based, and LLM-based. Human-centered approaches~\cite{human0, llmsurvey2} rely on domain experts to evaluate generated code based on their professional knowledge and programming experience. Although domain experts can offer accurate evaluations, these approaches are labour-intensive.
Metric-based approaches~\cite{bleu, codebleu, passk, passt, exec-acc} typically
require reference answers as ground truth or high-quality unit tests, which are generally difficult to create, maintain, and scale across diverse programming languages~\cite{llmsurvey, llmsurvey3}. 

\begin{figure}[!h]
	\centering
	\includegraphics[width=.5\textwidth]{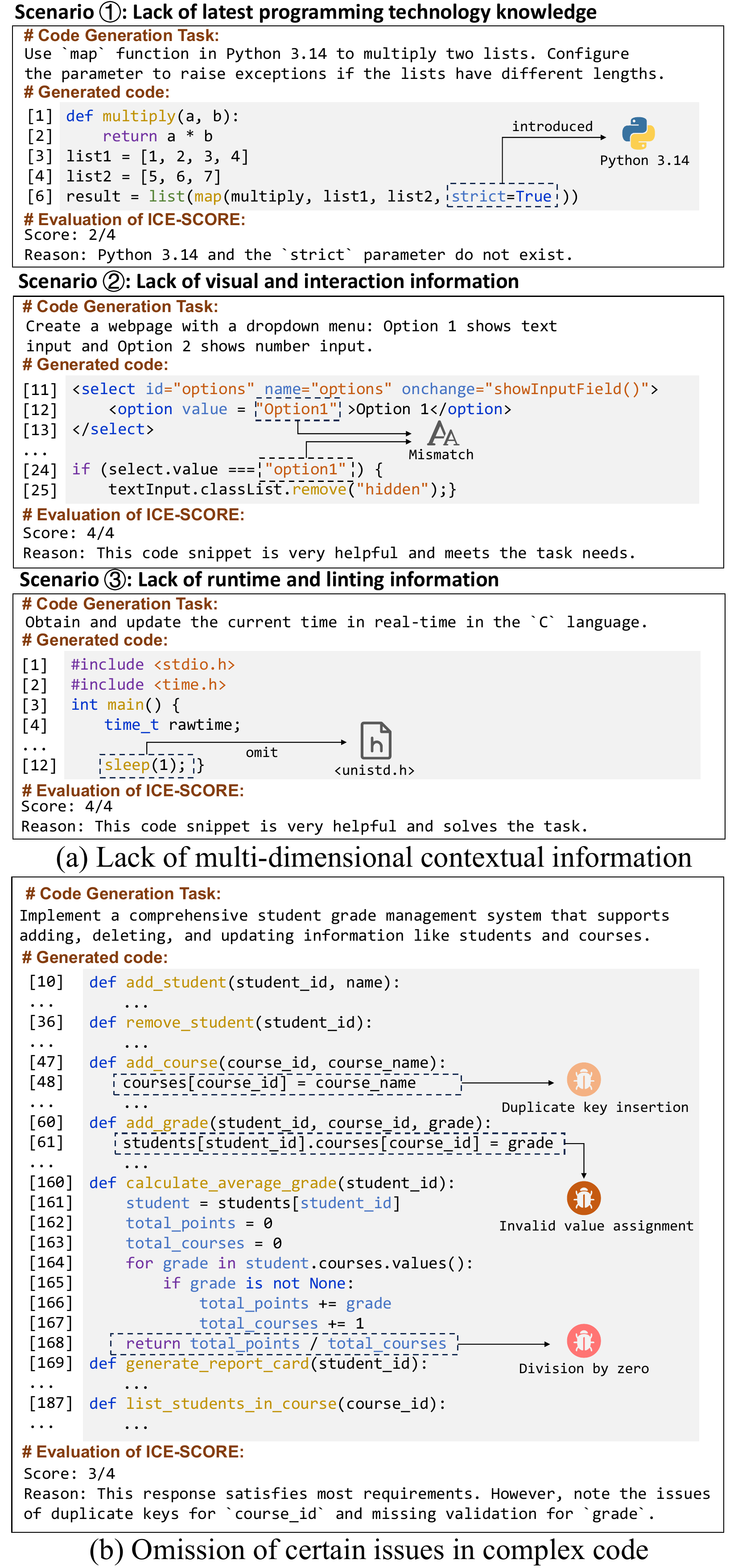}
\vspace{-1.5em}
    \caption{Examples for illustrating the limitations of LLM-based approaches for evaluating code generation. Each example includes the code generation task, the generated code, and the evaluation of ICE-SCORE. ICE-SCORE ratings range from 0 to 4, with higher scores indicating higher quality.
    }
\vspace{-1.5em}
\label{fig:challenges}
\end{figure}

Recently, LLM-based approaches~\cite{ice-score, codejudge} have gained increasing attention among researchers and practitioners. LLMs possess strong context-understanding and instruction-following capabilities, allowing them to perform more accurate evaluations. 
Compared to metric-based approaches, LLM-based approaches do not depend on specific benchmarks, eliminating the need to develop or identify suitable benchmarks, define precise evaluation metrics, or construct ground truth for tasks in the benchmark.
However, current LLM-based approaches still face limitations for complex code scenarios. To better demonstrate these limitations, we analyze the failure cases of ICE-SCORE~\cite{ice-score}, an advanced and representative LLM-based approach for evaluating code generation:
 
\textbf{(1) Lack of multi-dimensional contextual information:}
Complex code scenarios often encompass diverse requirements on runtime behavior, user interaction, and so on. Such requirements often rely on specific contextual information to be properly evaluated.
Figure~\ref{fig:challenges}(a) illustrates several scenarios where the existing LLM-based approaches fall short. \textbf{Scenario \textcircled{1}: Lack of latest programming technology knowledge.} When code generation tasks involve the latest programming technologies, these approaches fail to provide accurate evaluation due to a lack of relevant knowledge. For instance, a task requires using the ``\texttt{map}'' function in ``\texttt{Python 3.14}'' to multiply two lists, with the built-in parameter configured to check for length mismatches. 
The generated code correctly leverages the ``\texttt{map}'' function with a ``\texttt{strict}'' parameter (Line 6), which is introduced in ``\texttt{Python 3.14}'' to enforce length checking. However, ICE-SCORE assigns a score of 2, reasoning that both ``\texttt{Python 3.14}'' and the ``\texttt{strict}'' parameter do not exist. This instance highlights the inability of current LLM-based approaches to handle tasks involving the latest programming technologies.
\textbf{Scenario \textcircled{2}: Lack of visual and interaction information.} When involving front-end code, these approaches cannot view or interact with the graphical interface, leading to inaccurate evaluations. For instance, a task involves creating a webpage with a dropdown menu. In the generated code, the ``\texttt{select}'' component's ``\texttt{option}'' element assigns the value ``\texttt{Option1}'' (Line 12), while the conditional logic (Line 24) checks for ``\texttt{option1}''. This mismatch in case between ``\texttt{Option1}'' and ``\texttt{option1}'' prevents the intended interaction functionality.
However, because ICE-SCORE cannot simulate the graphical interface, it overlooks this issue and incorrectly evaluates the code as meeting the task requirements. This instance highlights that the lack of visual and interaction information may lead to flawed evaluations.
\textbf{Scenario \textcircled{3}: Lack of runtime and linting information.} Other information regarding code execution, syntax checking, and potential code issues is crucial for comprehensive evaluations. Without external tools, LLM-based approaches cannot access this information~\cite{llmsurvey}. For instance, a task requires generating code in ``\texttt{C}'' programming language to update the time in real-time. The generated code uses the ``\texttt{sleep}'' function (Line 12) but omits the necessary ``\texttt{<unistd.h>}'' header file. This omission would result in a compilation error. As ICE-SCORE cannot execute the code or leverage static analysis tools, it fails to detect this problem and incorrectly assigns a score of 4.
The examples indicate that the absence of multi-dimensional context tends to render LLM-based approaches inaccurate~\cite{llmsurvey}. 

\textbf{(2) Omission of certain issues in complex code:} 
Real-world \codetask often involve multiple requirements, resulting in generated code comprising numerous snippets with potential issues. Current LLM-based approaches struggle to fully comprehend and evaluate complex code in a fine-grained manner, as they typically require multiple reasoning steps and detailed evaluation of each task requirement and its corresponding code snippet~\cite{llmsurvey1, llmsurvey2}. Figure~\ref{fig:challenges}(b) illustrates a case of a student grade management system task. The generated code includes numerous functions, such as ``\texttt{add\_student}'', ``\texttt{remove\_student}'', and ``\texttt{calculate\_average\_grade}''.
ICE-SCORE scores the code 3, deeming it satisfies most task requirements and pinpointing issues such as duplicate keys for ``\texttt{course\_id}'' (Line 48) and missing validation for ``\texttt{grade}''. However, it overlooks another critical issue in the ``\texttt{calculate\_average\_grade}'' function (Line 160). Specifically, the return statement ``\texttt{return total\_points / total\_courses}'' (Line 168) does not handle division by zero when ``\texttt{total\_courses}'' is zero. This instance demonstrates that LLM-based approaches may overlook certain issues of complex code.

Besides, current LLM-based approaches solely provide an evaluation score without assessment details, which limits their usefulness in practical development workflows. As the automation of code generation evaluation is being integrated into the Continuous Integration/Continuous Deployment (CI/CD) pipeline of software development~\cite{llmsurvey}, real-time and comprehensive evaluation reports are essential for the rapid improvement of LLMs and better align with developers' preferences. 

\begin{figure*}[t]
	\centering
	\includegraphics[width=\textwidth]{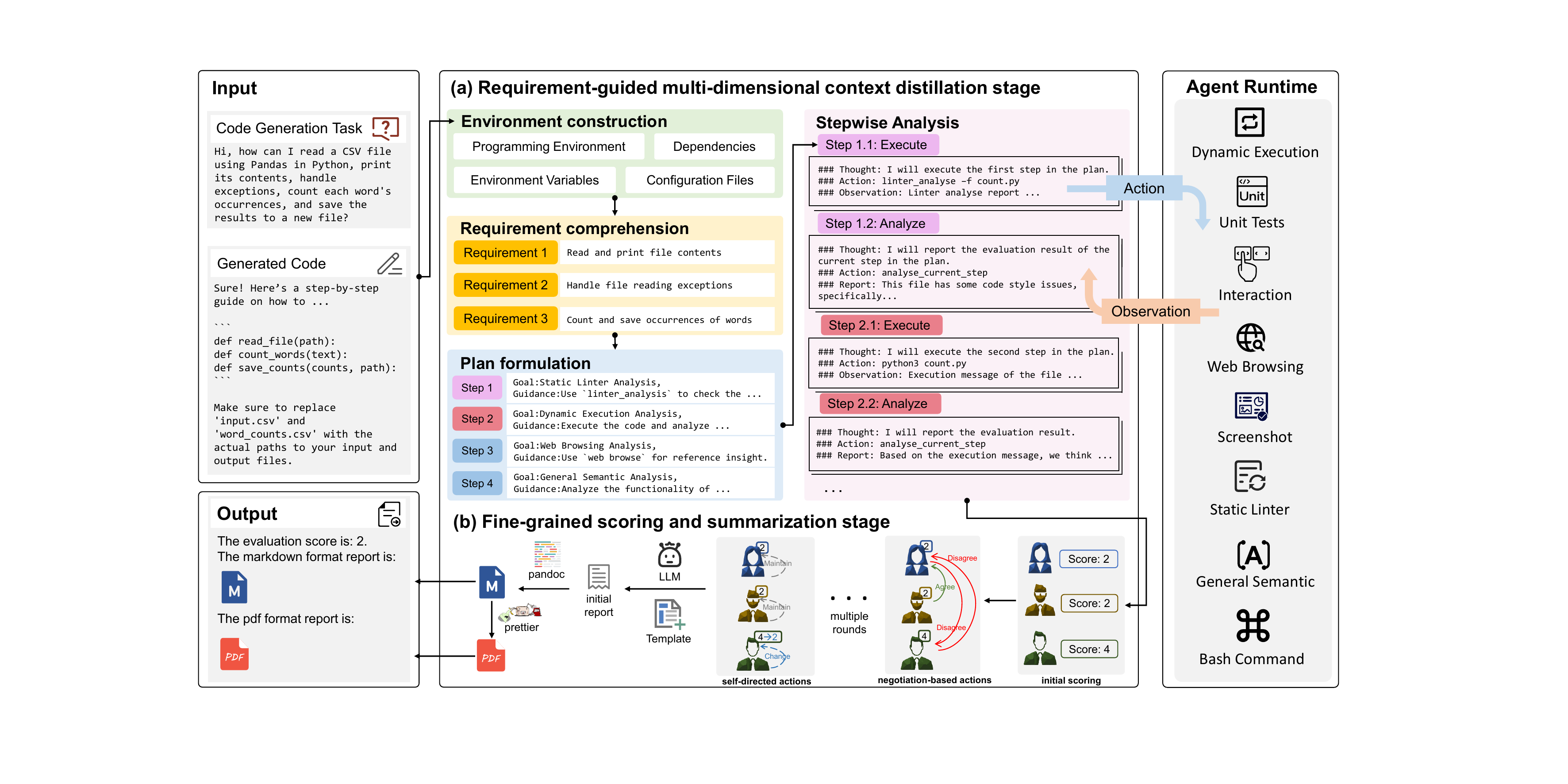}
    \vspace{-1.5em}
    \caption{The architecture of \framework. It consists of two main stages: (a) \moduleA for collecting contextual information based on the stepwise evaluation plan, and (b) \moduleB for generating evaluation scores and reports through negotiation with diverse viewpoints.}
\label{fig:framework}
\vspace{-1.5em}
\end{figure*}

Considering the limitations faced by the LLM-based approaches, we propose an agent-based evaluation framework for complex code generation, named \textbf{\framework}.  Agent-based frameworks can interact with the environment by invoking external tools and are capable of handling complex tasks~\cite{DBLP:journals/corr/abs-2405-03256, DBLP:journals/corr/abs-2312-13010, DBLP:journals/corr/abs-2403-14274}. Given a code generation task and the corresponding generated code, \framework evaluates the code through two stages, ultimately providing an evaluation score and evaluation report:
\textbf{(1) \moduleA}: We first formulate a detailed evaluation plan by decomposing task requirements, and then gather multi-dimensional contextual information for each requirement step by step.
\textbf{(2) \moduleB}: We define self-directed and negotiation-based strategies, enabling multiple judges to analyze complex code from fine-grained and diverse perspectives and reach consensus through discussion. Finally, we consolidate the assessment details and construct a structured evaluation report. 

We summarize the major contributions of this paper as follows:

(1) We propose the first agent-based evaluation framework for complex code generation, named \framework, to address the limitations of lacking multi-dimensional contextual information and complete analysis.

(2) We design a two-stage code evaluation pipeline for \framework: a \modulea for collecting contextual information, and a \moduleb for assessing generated code in a fine-grained manner. Finally, we provide comprehensive evaluation reports. 

(3) We conduct thorough experiments on \framework. The results demonstrate its effectiveness across different programming languages and coding scenarios. 

The remaining sections of this paper are organized as follows: 
Section~\ref{sec:evaluation} presents the architecture of \framework. Section~\ref{sec:experimental_setup} describes the experimental setup, including datasets, baselines, and experimental settings. 
Section~\ref{sec:experimental_result} reports and analyzes the experimental results.
Section~\ref{sec:discussion} further presents the effectiveness of \framework among different coding scenarios and programming languages, a case study of generated evaluation reports, performance on less complex benchmarks, and the threats to validity. Section~\ref{sec:related} introduces the background of code generation evaluation and LLM-based agents. 
Section~\ref{sec:conclusion} concludes the paper.

\section{Approach}
\label{sec:evaluation}
In this section, we describe the overall framework of \framework. 
As shown in Figure~\ref{fig:framework}, \framework consists of two stages: (1) \moduleA for gathering contextual information, and (2) \moduleB for scoring generated code through discussion between multiple judges. \framework utilizes LLMs as central control agents to conduct multi-turn interactions with the environment. During each interaction, the LLM agent responds with ``thought'' and ``action'', respectively, where ``thought'' represents its reasoning analysis of the environment feedback and ``action'' represents the command to be executed. As depicted in the ``Agent Runtime'' part of Figure~\ref{fig:framework}, the actions are executed within the environment, and the corresponding environment feedback is collected as ``observation'' to the LLM agent in the next interaction. 

\subsection{Requirement-guided Multi-dimensional Context Distillation Stage}
\label{stage1}
In this stage, \framework utilizes an LLM agent to gather multi-dimensional contextual information, including the latest programming technology knowledge, visual and interaction information, runtime and linting information, and so on. The LLM agent follows the paradigm of ``environment construction'', ``requirement comprehension'', ``plan formulation'', and ``stepwise analysis''. This paradigm mirrors the way humans address complex problems. 
Before solving complex problems, people typically decompose the requirements and formulate a plan, then proceed to analyze and solve the problem sequentially to ensure all requirements are satisfied, as recent studies~\cite{plan1, plan2} conclude that prompting LLMs to formulate plans before addressing problems is effective. Our designed paradigm is as follows:

\paragraph{\textbf{Environment Construction}} This phase aims to construct a complete and executable programming environment, serving as a necessary preparation for subsequent evaluations. 
Specifically, the agent configures the environment within a Docker container based on the code generation task and the generated code. 
The initial Docker container is configured with network settings and our custom external tools. 
Constructing the environment typically involves setting up the programming language runtime and installing the external dependencies. 
Figure~\ref{fig:framework} illustrates a task where the agent sets up the ``\texttt{Python}'' interpreter and installs the ``\texttt{Pandas}'' dependency for data processing. 
For more complex tasks, additional setup of configuration files and environment variables is required. 

\paragraph{\textbf{Requirement Comprehension}} 
This phase aims to enhance the agent's comprehension of the task requirements and prepare for formulating a thorough evaluation plan tailored to each requirement.
Specifically, the agent comprehends the code generation task and decomposes the task requirements. As illustrated in Figure~\ref{fig:framework}(a), the agent breaks down the code task into several detailed requirements: ``Read and print file contents'', ``Handle file reading exceptions'', and ``Count and save occurrences of words''. This decomposition enables the agent to better grasp complex tasks by dividing them into smaller, more manageable components.

\paragraph{\textbf{Plan Formulation}} 
This phase divides the evaluation of generated code into several steps, which facilitates the verification of each task requirement and reduces the reasoning burden on the agent.
Specifically, the agent formulates a detailed evaluation plan. Each step in the plan includes a brief \textbf{goal} (\ie, ``Goal'') and specific \textbf{guidance} (\ie, ``Guidance'') on how to achieve the goal. As illustrated in Figure~\ref{fig:framework}(a), the plan formulated by the agent includes goals such as ``Static Linter Analysis'', and ``Dynamic Execution Analysis'', along with corresponding guidance. The goal and guidance of each step in the plan collectively specify the action that the agent should take. For instance, the goal of the ``Static Linter Analysis'' is to ensure the generated code adheres to code syntax and style, and the guidance is to take the ``\texttt{linter\_analysis}'' action.  Below is a list of all possible actions during each step:

\begin{itemize}
    \item \textbf{Dynamic Execution:}
        This action checks for any compilation errors and analyzes whether the execution message aligns with the expected requirement. Possible commands include ``\texttt{python test.py}'' for running a Python script, ``\texttt{gcc -o output test.c \&\& ./output}'' for compiling and executing a C program, and so on.

    \item \textbf{Static Linter:} 
        This action checks for syntax errors, code issues, and style inconsistencies. It helps identify potential problems in the generated code without execution. We design the command ``\texttt{linter\_analysis -f `path'}'' to check a file or directory for syntax compliance. This command automatically selects appropriate static analysis tools for the given file path and returns a structured analysis report.

    \item \textbf{Unit Tests:}
        This action involves writing and executing unit tests, ensuring the generated code meets specific requirements and behaves as expected under various conditions. These unit tests support requirement analysis by providing different inputs and validating the outputs through assertions. For instance, consider a function that processes a list of integers and returns a list of their squares, but only for even numbers. A unit test might provide the input ``\texttt{[1, 2, 3, 4]}'' and utilize the assertion ``\texttt{assert process\_list([1, 2, 3, 4]) == [4, 16]}'' to verify that the output is as expected. 

    \item \textbf{Screenshot:}
        This action involves capturing a screenshot of the front-end code to support requirement analysis through visualization. We design the command ``\texttt{screenshot\_analysis -f `path' -q `query'}'', which renders the specified file as a screenshot and generates an analysis report utilizing a multimodal LLM. The agent can also query the screenshot to extract relevant visual information, such as UI elements and layout.

    \item \textbf{Interaction:}
        This action involves simulating user interactions with the front-end code to evaluate the interaction requirements. We extend the ``\texttt{screenshot\_analysis}'' command with the ``\texttt{-a `actions'}'' parameter to interact with the graphical interface before capturing the screenshot. Possible interactions include clicking elements, filling input fields, hovering over elements, and scrolling. 

    \item \textbf{Web Browsing:}
        This action involves retrieving additional information from the website, including knowledge of the latest and specialized programming technologies. We design the command ``\texttt{web\_browse -q `query'}'' to search the website according to the query. 
        
    \item \textbf{General Semantic:}
         Considering the LLM's powerful contextual understanding capabilities, the agent can also evaluate whether the generated code satisfies requirements related to complexity, security, and robustness, without invoking external tools.

    \item \textbf{Bash Command:}
        In addition to the actions designed specifically for code generation, the agent also needs to call bash commands during the evaluation process to perform steps such as writing code, writing data, viewing files, and navigating the file system.
\end{itemize}

\paragraph{\textbf{Stepwise Analysis}} 
In this phase, the agent follows the evaluation plan by executing the specified action at each step and analyzing the corresponding execution results. The agent alternates between executing and analyzing during interactions, with its current state indicated as ``\texttt{Execute State}'' or ``\texttt{Analyze State}''.
\begin{itemize}
    \item \textbf{\texttt{Execute State:}}
        In this state, the agent takes the action specified in the current step.
        The action is executed within the constructed Docker container, and the corresponding execution results are collected as the environment feedback (\ie, ``observation'') to the agent. Besides, the following hint is given as additional environment feedback to facilitate the evaluation process:
\end{itemize}
\begin{tcolorbox}[width=.49\textwidth, colframe=purple!20!white, colback=purple!5!white, boxrule=0.2mm, arc=0.5mm, fonttitle=\small, coltitle=black]
    \small
    \textbf{Hint:} \textit{Once you have executed the actions in the current step, you should analyze and report the execution results in the next interaction. If the step is not yet finished, please continue working on it.}
\end{tcolorbox}
\begin{itemize}
    \item \textbf{\texttt{Analyze State:}}
        In this state, the agent takes the action ``\texttt{analyze\_current\_step}'' to analyze and report the execution results. We predefine different analysis report templates tailored to the execution results of different actions. 
        Similarly, the following hint is provided as additional environment feedback:
\end{itemize}
\begin{tcolorbox}[width=.49\textwidth, colframe=purple!20!white, colback=purple!5!white, boxrule=0.2mm, arc=0.5mm, fonttitle=\small, coltitle=black]
    \small
    \textbf{Hint:} \textit{You have submitted the analysis report of the current step. Please review the steps you have already completed and proceed to the next step.}
\end{tcolorbox}

Besides, the agent can adjust the evaluation plan in real time based on the actual evaluation process to enhance flexibility. 
Finally, the analysis reports from each step, together with the code generation task, generated code, and decomposed requirements, are collected as input for the next stage.

\subsection{Fine-grained Scoring And Summarization Stage}

In this stage, \framework employs multiple LLM agents to act as judges, scoring the generated code through collaborative discussion.
Since a single judge may potentially overlook issues in complex code or make misjudgements, involving multiple judges enables more fine-grained evaluation by incorporating diverse viewpoints and facilitating consensus. This negotiation strategy can improve the thoroughness and reliability of the evaluation results, as different judges focus on different issues and aspects of the generated code.

\subsubsection{Definition of Judges}

Let \( A_1, A_2, \ldots, A_n \) denote the \( n \) LLM agents acting as individual judges. Each judge \( A_i \) assigns a score to the generated code based on the decomposed requirements, stepwise analysis reports from Section~\ref{stage1}, and the evaluation criteria. The evaluation criteria include aspects of correctness, functionality, and clarity. Below, we present an example of the clarity aspect and its corresponding
evaluation criteria, while the remaining aspects and criteria are available in our repository:
\begin{tcolorbox}[width=.49\textwidth, colframe=blue!12!white, colback=blue!5!white, boxrule=0.2mm, arc=0.5mm, fonttitle=\small, coltitle=black]
    \small
    \textbf{Clarity:} \textit{Is the generated code explained in an easy-to-understand manner, logically clear, and unambiguous?}
 
    \textbf{- 0/4:} \textit{Mostly unclear or confusing; very difficult to understand.}
    
    \textbf{- 1/4:} \textit{Significant ambiguity, partially understandable; contains several confusing parts.}
    
    \textbf{- 2/4:} \textit{Some ambiguity, but mostly understandable; contains some unclear parts.}
    
    \textbf{- 3/4:} \textit{Mostly clear, slight ambiguity; generally easy to understand with minor issues.}
    
    \textbf{- 4/4:} \textit{Very clear, no ambiguity, well-organized; completely easy to understand.}
\end{tcolorbox}

Each judge provides an evaluation score \( S_i \) and corresponding scoring reason \( R_i \). Formally, each judge \( A_i \) is represented as a tuple:
\begin{equation}
A_i = (S_i, R_i)
\end{equation}
where \( S_i \) and \( R_i \) denote the score and reason provided by the judge \( A_i \), respectively.

\subsubsection{Negotiation and Scoring}
\label{stage2_2}

After each judge \( A_i \) has provided their initial score \( S_i^0 \) and reason \( R_i^0 \), these evaluations are shared among all other judges, ensuring that each judge is aware of their peers’ assessments. This process can be formally expressed as:
\begin{equation}
\forall i, j \in \{1, 2, \ldots, n\}, \quad (S_i^0, R_i^0) \rightarrow A_j
\end{equation}

Then these judges engage in multiple rounds of negotiation to reach a consensus. Let \( k \) denote the round number. In each round \( k \), each judge \( A_i \) may take one or more actions \( Action_i^k \), which can be categorized into two types: self-directed actions and negotiation-based actions.

Self-directed actions allow a judge to manage their own evaluation, including maintaining the current score \( S_i \) and the reason \( R_i^0 \)\texttt{(Maintain)}, changing the score \( S_i \) and the reason \( R_i^0 \) \texttt{(Change)}, or withdrawing a previously stated opinion from round \( k \) in light of possible opinion changes \texttt{(Withdraw)}. Formally, these actions are defined as:
\begin{equation}
\text{Actions}_{\text{self-directed}} = 
\left\{
\begin{aligned}
&\text{Maintain}(E) \\
&\text{Change}(S_i, R_i, E) \\
&\text{Withdraw}(k, E)
\end{aligned}
\right\}
\end{equation}

Negotiation-based actions, on the other hand, facilitate consensus-building among judges. These actions correspond to three forms of negotiation, including explicitly agreeing with another judge's score and reason \texttt{(Agree)}, requesting clarification \texttt{(Query)}, or expressing disagreement \texttt{(Disagree)}. Formally, these actions are defined as:
\begin{equation}
\text{Actions}_{\text{negotiation-based}} = 
\left\{
\begin{aligned}
&\text{Agree}(A_j, E) \\
&\text{Query}(A_j, E) \\
&\text{Disagree}(A_j, E)
\end{aligned}
\right\}
\end{equation}

Hence, the \( Action_i^k \) taken by the judge \( A_i \) in round \( k \) is given by:
\begin{equation}
\text{Action}_i^k \subseteq \left\{ \text{Actions}_{\text{self-directed}} \cup \text{Actions}_{\text{negotiation-based}} \right\}
\end{equation}

These judges are required to provide an explanation (\ie, \( E \)) while taking different actions. After each round \( k \), the current score and reason of each judge, as well as the actions taken, are shared among all judges:
\begin{equation}
\forall i, j \in \{1, 2, \ldots, n\}, \quad (S_i^k, R_i^k, Action_i^k) \rightarrow A_j
\end{equation}

The negotiation ends when either the number of rounds is exceeded or all judges reach a consensus on the evaluation score. The final evaluation score is calculated as the average of the scores provided by all judges:
\begin{equation}
\text{Evaluation Score} = \frac{1}{n} \sum_{i=1}^{n} S_i
\end{equation}

\subsubsection{Summarization and Evaluation Report Generation}

We obtain the environment configuration, task requirements, and stepwise analysis reports in Section \ref{stage1} and the final evaluation score in Section \ref{stage2_2}. Besides, \framework leverages another LLM to conclude the overall evaluation results, which consist of the evaluation reasons and the optimization suggestions provided by multiple judges after negotiation. Formally, this process can be represented as: 
\begin{align}
\text{Evaluation Score}, \{(S_i, R_i) \mid i = 1, 2, \ldots, n\} 
\xrightarrow{\quad} \notag \\
\text{Evaluation Reason}, \text{Optimization Suggestion}
\end{align}

Then the agent generates the initial evaluation report in Markdown format. To enhance readability, we standardize and beautify the initial evaluation report.  We utilize Prettier~\cite{prettier} to check for formatting issues and maintain a consistent style. Besides, we employ Pandoc~\cite{pandoc} to convert the Markdown report to PDF format.

\begin{table}[t]
\centering
\caption{Statistics of the constructed benchmark.}
\setlength{\tabcolsep}{8mm}
\renewcommand{\arraystretch}{0.9}
\label{tab:statistics}
\begin{tabular}{lr}
\toprule
\rowcolor{gray!25} \textbf{Statistics} & \textbf{Number} \\ \midrule
\textbf{Samples} & 363 \\
\textbf{Code Generation Tasks} & 121 \\
\textbf{Coding Scenarios} & 37 \\
\textbf{Programming Languages} & 23 \\
\midrule
\textbf{Average String Length} & \\
\quad Code Generation Task & 1,906  \\
\quad LLM-generated Response & 2,650 \\
\bottomrule
\end{tabular}
\vspace{-1em}
\end{table}

\section{Experimental Setup}
\label{sec:experimental_setup}
In this section, we evaluate the \framework and aim to answer the following research questions (RQs):

\begin{enumerate}[label=\bfseries RQ\arabic*:,leftmargin=.5in]
    
    \item How does \framework perform in evaluating code generation compared with different methods?
    \item What is the influence of different stages on the performance of \framework?
    \item How do different hyper-parameters impact the performance of \framework?
\end{enumerate}

\subsection{Datasets}

CodeArena~\cite{codearena} is a human-curated benchmark of 397 high-quality code generation tasks, covering various coding scenarios and programming languages. It simulates the complexity and diversity of real-world tasks, with each task classified as ``easy'', ``medium'', or ``hard'' based on complexity. To evaluate \framework's performance across diverse and complex tasks, we filter CodeArena, generate responses using different LLMs, and manually score the responses. The filtering, response generation, and manual scoring process is as follows:

    (1) Filtering:
        We first retain tasks classified as ``hard''. Then, we filter out tasks that depend on specific software or platforms, such as those requiring MATLAB or Verilog.

    (2) Response Generation:
        We generate responses for each code generation task utilizing three popular LLMs, including GPT-3.5-turbo~\cite{gpt35}, Claude-3.5-Sonnet~\cite{claude}, and GPT-4o~\cite{gpt4O}.

    (3) Manual Scoring:
    Each code generation task and corresponding LLM-generated response is scored by two experienced experts, each with over five years of expertise in the relevant programming languages. Scoring is conducted on a scale from 0 to 4, where higher scores indicate better response quality, following the widely-used evaluation criteria~\cite{human0}. Due to page limitations, the detailed evaluation criteria are available in our repository. The Kappa coefficient between the two experts exceeds 80\%, indicating a high level of agreement. In case of scoring discrepancies, a third expert is introduced to adjudicate. All experts remain unaware of the identity of the LLM that generates each response.

The statistics of our constructed benchmark are illustrated in Table~\ref{tab:statistics}. The benchmark covers 37 coding scenarios and 23 programming languages, enabling thorough assessment of \framework across diverse tasks. Each sample is represented as a triplet: (code generation task, LLM-generated response, human-annotated score).



\begin{table*}[h!]
\centering
\caption{Comparison results between \framework and baseline methods.}
\setlength{\tabcolsep}{2.3mm}
\begin{tabular}{lccccccccc|ccc}
\toprule
\rowcolor{gray!25}
\textbf{Models} & \multicolumn{3}{c}{GPT-3.5-turbo} & \multicolumn{3}{c}{Claude-3.5-Sonnet} & \multicolumn{3}{c}{GPT-4o} & \multicolumn{3}{|c}{\textbf{AVG}} \\
\cmidrule{1-4} \cmidrule{5-7} \cmidrule{8-10} \cmidrule{11-13}
\rowcolor{gray!25}
\textbf{Metrics} & $r_p$ & $r_s$ & $\tau$ & $r_p$ & $r_s$ & $\tau$ & $r_p$ & $r_s$ & $\tau$ & $r_p$ & $r_s$ & $\tau$ \\
\midrule
\textbf{VANILLA} & 0.083 & 0.129 & 0.117 & 0.033 & 0.098 & 0.094 & 0.010 & 0.047 & 0.045 & 0.042 & 0.091 & 0.085 \\
\textbf{ICE-SCORE} & 0.100 & 0.177 & 0.157 & 0.029 & 0.096 & 0.092 & -0.011 & 0.054 & 0.051 & 0.039 & 0.109 & 0.100 \\
\textbf{CODEJUDGE} & 0.130 & 0.130 & 0.120 & 0.097 & 0.074 & 0.072 & 0.025 & 0.078 & 0.076 & 0.084 & 0.094 & 0.089\\ \midrule
\textbf{\framework} & \textbf{0.325} & \textbf{0.286} & \textbf{0.247} & \textbf{0.317} & \textbf{0.278} & \textbf{0.262} & \textbf{0.262} & \textbf{0.253} & \textbf{0.214} & \textbf{0.301} & \textbf{0.272} & \textbf{0.241} \\
\bottomrule
\end{tabular}
\label{tab:comparison}
\end{table*}

\subsection{Comparison Baselines}

To evaluate the performance of \framework, we compare the state-of-the-art LLM-based approaches. Following the previous work~\cite{codejudge}, we also introduce a vanilla approach as the baseline method.

\begin{itemize}
    \item \textbf{VANILLA} directly prompts LLMs to determine the correctness and helpfulness of the response. 
    \item \textbf{ICE-Score~\cite{ice-score}} utilizes assessment criteria and an evaluation step template to evaluate the usefulness and functional correctness of the response. 
    \item \textbf{CODEJUDGE~\cite{codejudge}}  guides LLMs in performing ``slow thinking'' to arrive at an in-depth and reliable evaluation of semantic correctness.
\end{itemize}

\subsection{Evaluation Metrics}

We follow best practices in natural language generation evaluation and utilize Pearson ($r_p$), Spearman ($r_s$), and Kendall-Tau ($\tau$) coefficients to measure the correlation between evaluation scores assigned by different approaches and the ground truth (\ie, scores annotated by humans).  

\subsection{Implementation Details}
\framework and the baseline methods are provided access to GPT-4o. For the \modulea, we limit the maximum number of interactions to 40 and set the sampling temperature to 0.2. For the \moduleb, we set the sampling temperature to 0.7 and the number of judges to 3, and allow up to 4 rounds of negotiation. For the baseline methods, we try our best to reproduce them from publicly available source code and papers, and use the same hyper-parameter settings. For all correlation metrics, we use the implementation from SciPy~\cite{scipy} and call these APIs with the default settings.



 
\section{Experimental Results}
\label{sec:experimental_result}
\subsection{RQ1: Effectiveness of \framework in Evaluating Code Generation}
To answer RQ1, we conduct a comprehensive analysis against three baseline methods. The experimental results are shown in Table~\ref{tab:comparison}. 
 
 \textbf{\framework exhibits superior performance compared with the baseline methods.} As shown in Table~\ref{tab:comparison}, we observe that \framework consistently outperforms all the baseline methods across three LLMs (\ie, responses generated by GPT-3.5-turbo, Claude-3.5-Sonnet, and GPT-4o). 
Overall, \framework achieves average scores of 0.301, 0.272, and 0.241 on $r_p$, $r_s$, and $\tau$, respectively, outperforming the best baseline methods with improvements of 0.217, 0.163, and 0.141. These improvements are due to \framework's ability to collect multi-dimensional contextual information and provide the evaluation score through fine-grained negotiation.

\textbf{Current advanced LLM-based approaches show minimal difference compared to VANILLA.} 
We observe that state-of-the-art LLM-based approaches, ICE-SCORE and CODEJUDGE, exhibit little difference compared to VANILLA. Specifically, the average scores of VANILLA on $r_p$, $r_s$, and $\tau$ are 0.042, 0.091, and 0.085, respectively, while ICE-SCORE achieves 0.039, 0.109, and 0.100, and CODEJUDGE achieves 0.084, 0.094, and 0.089, respectively. These results indicate that the differences are marginal. Besides, VANILLA outperforms the other two methods in $r_s$ and $\tau$ when evaluating responses generated by Claude-3.5-Sonnet. These results further demonstrate that current LLM-based methods struggle to comprehend complex code without contextual information, which is essential for accurate and comprehensive code evaluation.

\begin{tcolorbox}
 \textbf{Answer to RQ1:} \framework achieves the best performance on evaluating code generation, exceeding the best baseline method by 0.217, 0.163, and 0.141 on $r_p$, $r_s$, and $\tau$, respectively.
\end{tcolorbox}

\subsection{RQ2: Effectiveness of Different Stages in \framework}

\begin{table*}[h!]
\centering
\setlength{\tabcolsep}{2.3mm}
\caption{Impact of different stages on the performance of \framework.}
\begin{tabular}{lccccccccc|ccc}
\toprule \rowcolor{gray!25}
\textbf{Models} & \multicolumn{3}{c}{GPT-3.5-turbo} & \multicolumn{3}{c}{Claude-3.5-Sonnet} & \multicolumn{3}{c}{GPT-4o} & \multicolumn{3}{|c}{\textbf{AVG}} \\
\cmidrule{1-4} \cmidrule{5-7} \cmidrule{8-10} \cmidrule{11-13}
\rowcolor{gray!25}
\textbf{Metrics} & $r_p$ & $r_s$ & $\tau$ & $r_p$ & $r_s$ & $\tau$ & $r_p$ & $r_s$ & $\tau$ & $r_p$ & $r_s$ & $\tau$ \\
\midrule
\textbf{w/o RMCD}  & 0.278 & 0.228 & 0.200 & 0.250 & 0.219 & 0.198 & 0.182 & 0.171 & 0.147 & 0.237 & 0.206 & 0.182 \\
 $\hookrightarrow$ w/o RT  & 0.229 & 0.208 & 0.185 & 0.233 & 0.219 & 0.193 & 0.152 & 0.153 & 0.133 & 0.205 & 0.193 & 0.170\\
 $\hookrightarrow$ w/o UI/UX  & 0.305 & 0.258 & 0.217 & 0.302  & 0.242 & 0.216 & 0.198 & 0.214 & 0.169 & 0.268 & 0.238 & 0.201 \\
 $\hookrightarrow$ w/o ST  & 0.290 & 0.286 & 0.234 & 0.280 & 0.179 & 0.157 & 0.176 & 0.161 & 0.136 & 0.249 & 0.209 & 0.176\\ \midrule
\textbf{w/o FSAS} & 0.282 & 0.233 & 0.204 & 0.262 & 0.250 & 0.227 & 0.201 & 0.189 & 0.172 & 0.248 & 0.224 & 0.201 \\ \midrule

\textbf{\framework} & \textbf{0.325} & \textbf{0.286} & \textbf{0.247} & \textbf{0.317} & \textbf{0.278} & \textbf{0.262} & \textbf{0.262} & \textbf{0.253} & \textbf{0.214} & \textbf{0.301} & \textbf{0.272} & \textbf{0.241}\\
\bottomrule
\end{tabular}
\label{tab:ablation}
\end{table*}


To answer RQ2, we explore the effectiveness of different stages on the performance of \framework.
\subsubsection{Requirement-guided Multi-dimensional Context Distillation Stage} To understand the impact of this stage, we deploy a variant of \framework without the \modulea (\ie, w/o RMCD). 
The variant follows a completely free evaluation process instead of following our designed paradigm of ``environment construction'', ``requirement comprehension'', ``plan formulation'', and ``stepwise analysis''. 
As shown in Table~\ref{tab:ablation}, the addition of the RMCD stage achieves higher performance on $r_p$, $r_s$, and $\tau$ across all three LLMs. 
Specifically, adding the RMCD stage improves performance by an average of 27.0\%, 32.0\%, and 32.4\% across the three metrics, respectively.
The RMCD stage excels in understanding diverse requirements and formulating detailed evaluation plans, thereby ensuring the thorough collection of multi-dimensional context.

\subsubsection{Fine-grained Scoring And Summarization Stage}
To explore the contribution of this stage, we also construct a variant of \framework without the \moduleb (\ie, w/o FSAS).
This variant scores the LLM-generated response multiple times and takes the average instead of a discussion between multiple judges. As shown in Table~\ref{tab:ablation}, the addition of the FSAS stage achieves higher performance on $r_p$, $r_s$, and $\tau$ across all three LLMs. 
Specifically, adding the FSAS stage improves performance by 21.4\%, 21.4\%, and 19.9\% on average across the three metrics, respectively.
The FSAS stage leverages multiple judges for negotiation, mitigating the inaccuracy of comprehending complex code and integrating diverse opinions.

\subsubsection{Different Information in Requirement-guided Multi-dimensional Context Distillation Stage}
To further investigate the influence of different information in the \modulea, we design three variants by disabling different information sources: 
test execution (runtime information, \ie, RT), 
screenshot and interaction (UI/UX information), 
and web browsing and syntax linters (static information, \ie, ST). 
As shown in Table~\ref{tab:ablation}, different contextual information contributes to improvements in evaluation accuracy. 
Incorporating runtime information yields the most notable gains, improving $r_p$, $r_s$, and $\tau$ by 46.8\%, 40.9\%, and 41.8\%, respectively, highlighting the strong impact of runtime information in capturing execution anomalies and boundary conditions. UI/UX information provides a multimodal understanding, leading to moderate improvements of 12.3\%, 14.3\%, and 19.9\%, respectively. Other static information from web browsing and syntax linters helps identify potential code issues and provides external context, resulting in gains of 20.9\%, 30.1\%, and 36.9\%, respectively.
Overall, these results demonstrate that different types of information each play a complementary role in enhancing evaluation performance.

\begin{tcolorbox}
 \textbf{Answer to RQ2:} Both RMCD and FSAS stages can improve the performance of \framework. The RMCD stage boosts $r_p$, $r_s$, and $\tau$ by 27.0\%, 32.0\%, and 32.4\%, respectively, while the FSAS stage enhances \framework by 21.4\%, 21.4\%, and 19.9\%, respectively. Besides, different information in the RMCD is essential to accurate evaluation.
\end{tcolorbox}

\subsection{RQ3: Influence of Hyper-parameters on the Performance of \framework}

To answer RQ3, we explore the impact of different hyper-parameters, including the number of judges and the maximum number of rounds during the \moduleb.

\subsubsection{Number of Judges}

Figure~\ref{fig:hyper}(a) shows the performance of \framework with different numbers of judges. As the number of judges increases from 2 to 3, the performance improves notably, with $r_p$, $r_s$, and $\tau$ increasing from 0.321, 0.289, and 0.261 to 0.365, 0.353, and 0.303, respectively. When the number increases to 5, the performance stabilizes, indicating that further increases have a limited impact. This suggests that three judges are sufficient to provide a comprehensive evaluation, and additional judges struggle to offer new insights. Considering the balance between resource consumption and performance, we choose 3 as the optimal number of judges.

\subsubsection{Maximum Number of Rounds}

\begin{figure}[t]
	\centering
	\includegraphics[width=.5\textwidth]{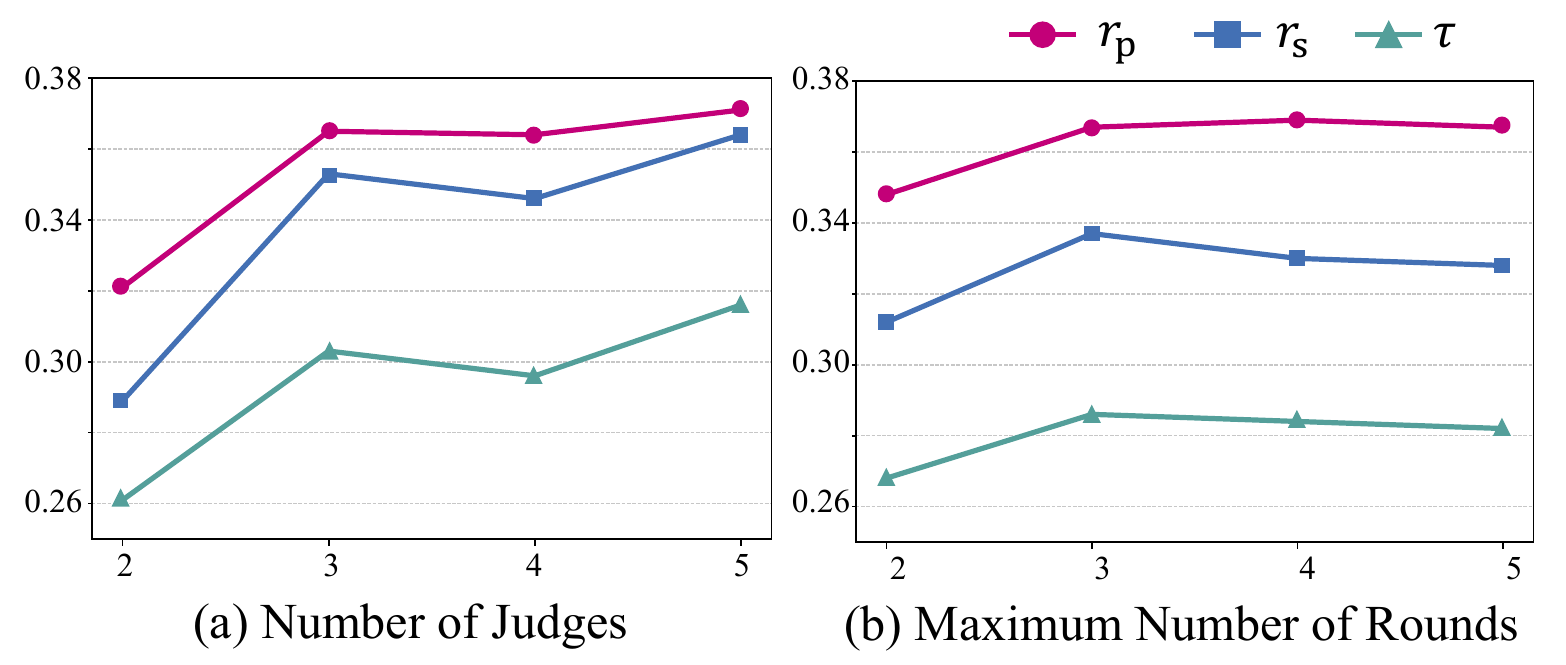}
    \caption{The influence of the number of judges and maximum number of rounds on \framework. The horizontal axis represents the number of judges or the maximum number of rounds.}
\label{fig:hyper}
\vspace{-1.5em}
\end{figure}

Similarly, as shown in Figure~\ref{fig:hyper}(b), increasing the maximum number of rounds from 2 to 3 leads to steady performance improvements, with $r_p$, $r_s$, and $\tau$ rising from 0.348, 0.312, and 0.268 to 0.367, 0.337, and 0.286, respectively. This demonstrates that negotiation allows judges to effectively share insights and mitigate biases in understanding complex code, thereby refining their evaluations. However, as the number of rounds increases from 3 to 5, the performance tends to fluctuate slightly, suggesting that consensus has already been reached and additional rounds bring little to no benefit. To balance resource consumption and performance, we select 4 as the optimal number of rounds.

\begin{tcolorbox}
 \textbf{Answer to RQ3:} 
 The performance of \framework is influenced by the number of judges and the maximum number of rounds. Our default settings of 3 judges and 4 rounds yield optimal results.
\end{tcolorbox}






 
\section{Discussion}
\label{sec:discussion}
\subsection{Performance on Different Coding Scenarios and Programming Languages}

\begin{figure}[t]
	\centering
	\includegraphics[width=.5\textwidth]{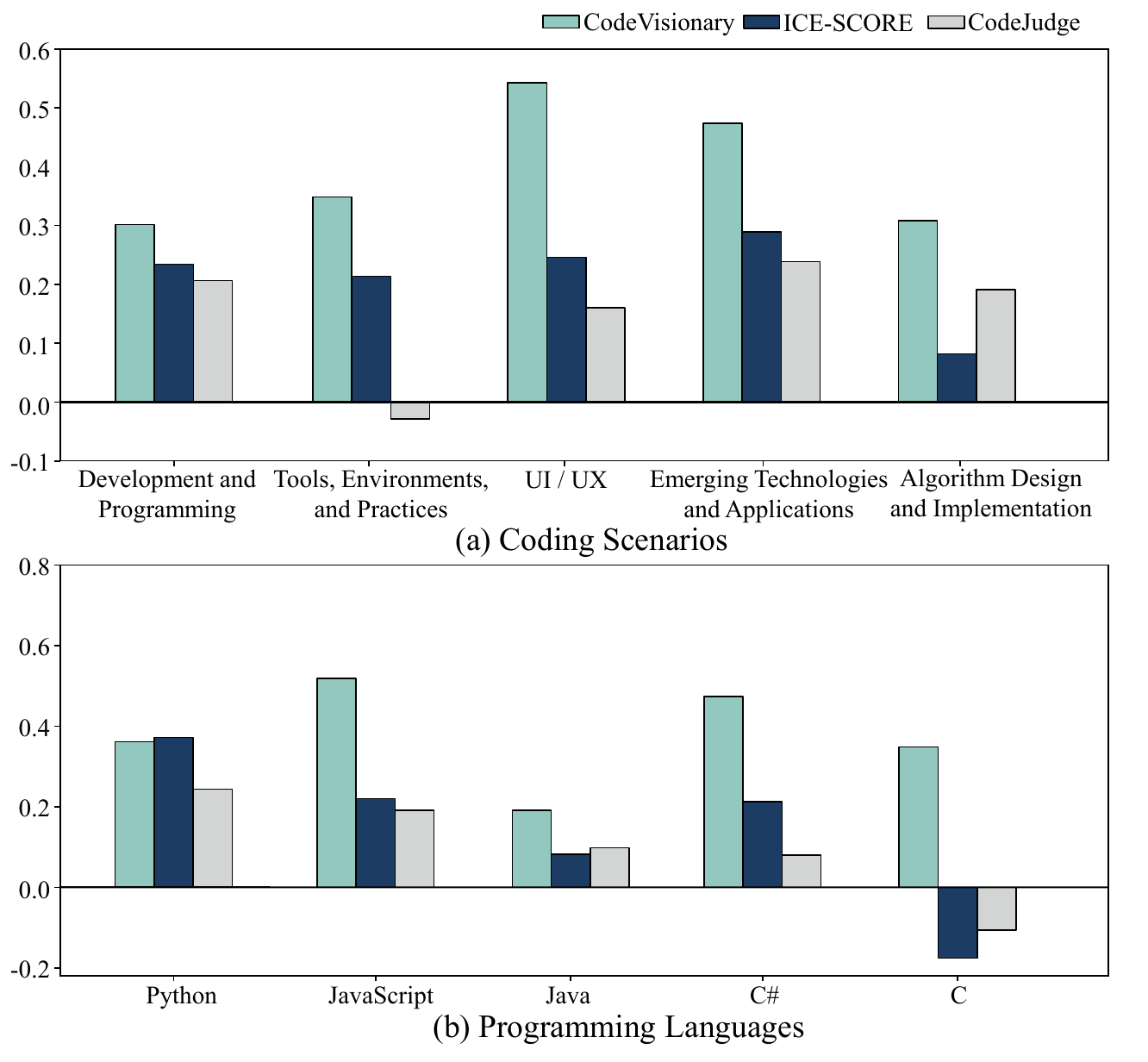}
    \caption{Performance of \framework and baseline methods across different coding scenarios and programming languages, measured by $r_s$.}
\label{fig:display}
\end{figure}

\begin{table}[t]
\centering
\setlength{\tabcolsep}{6pt} 
\renewcommand{\arraystretch}{0.9}
\caption{Counts and percentages of different actions invoked by \framework across all instances.}
\begin{tabular}{lccr}
\toprule
\rowcolor{gray!25} \textbf{Action} & \textbf{\# Total} & \textbf{\# Per Instance} & \textbf{\%} \\
\midrule
Dynamic Execution & 210 & 0.58 & 9.45 \\
Static Linter     & 366 & 1.01 & 16.46 \\
Unit Tests        & 84  & 0.23 & 3.78 \\
Screenshot        & 76  & 0.21 & 3.42 \\
Interaction       & 53  & 0.15 & 2.38 \\
Web Browsing      & 325 & 0.90 & 14.62 \\
General Semantic  & 652 & 1.80 & 29.33 \\
Bash Command      & 343 & 0.94 & 15.43 \\
Other      & 114 & 0.31 & 5.13 \\
\midrule
All     & 2,223 & 6.12 & 100.00 \\
\bottomrule
\end{tabular}
\label{tab:action_counts}
\vspace{-1em}
\end{table}

Figure~\ref{fig:display} illustrates the performance of \framework and baseline methods across mainstream coding scenarios and programming languages. As shown in Figure~\ref{fig:display}(a), \framework generally surpasses baseline methods in different coding scenarios. In particular, CodeVisionary improves the $r_s$ in ``UI/UX'', ``Emerging Technologies and Applications'', and ``Algorithm Design and Implementation'' from 0.246 to 0.543, 0.289 to 0.474, and 0.191 to 0.308, respectively. Table~\ref{tab:action_counts} presents the counts and percentages of various actions invoked by \framework. In the ``UI/UX'' scenario, \framework employs ``Screenshot'' (3.42\%) and ``Interaction'' (2.38\%) to perform multimodal assessment of front-end code. For the ``Emerging Technologies and Applications'' scenario, it can leverage ``Web Browsing'' (14.62\%) to acquire the latest or specific programming knowledge. In the ``Algorithm Design and Implementation'' scenario, \framework can utilize ``Dynamic Execution'' (9.45\%) and ``Unit Tests'' (3.78\%) to test boundary cases through code execution.

Similarly, Figure~\ref{fig:display}(b) shows that \framework generally outperforms baseline methods across different programming languages. It achieves much higher $r_s$ in ``JavaScript'', ``C\#'', and ``C'', increasing from 0.220 to 0.519, 0.213 to 0.473, and -0.106 to 0.348, respectively. The strong performance in ``JavaScript'' can be attributed to the domain information gained from ``Screenshot'' and ``Interaction''. Regarding ``C'', the negative coefficient of baseline methods highlights the importance of execution information. 
Overall, these results demonstrate that \framework effectively adapts across coding scenarios and programming languages.

\subsection{Case Study of Evaluation Reports}

\begin{figure}[t]
	\centering
	\includegraphics[width=.5\textwidth]{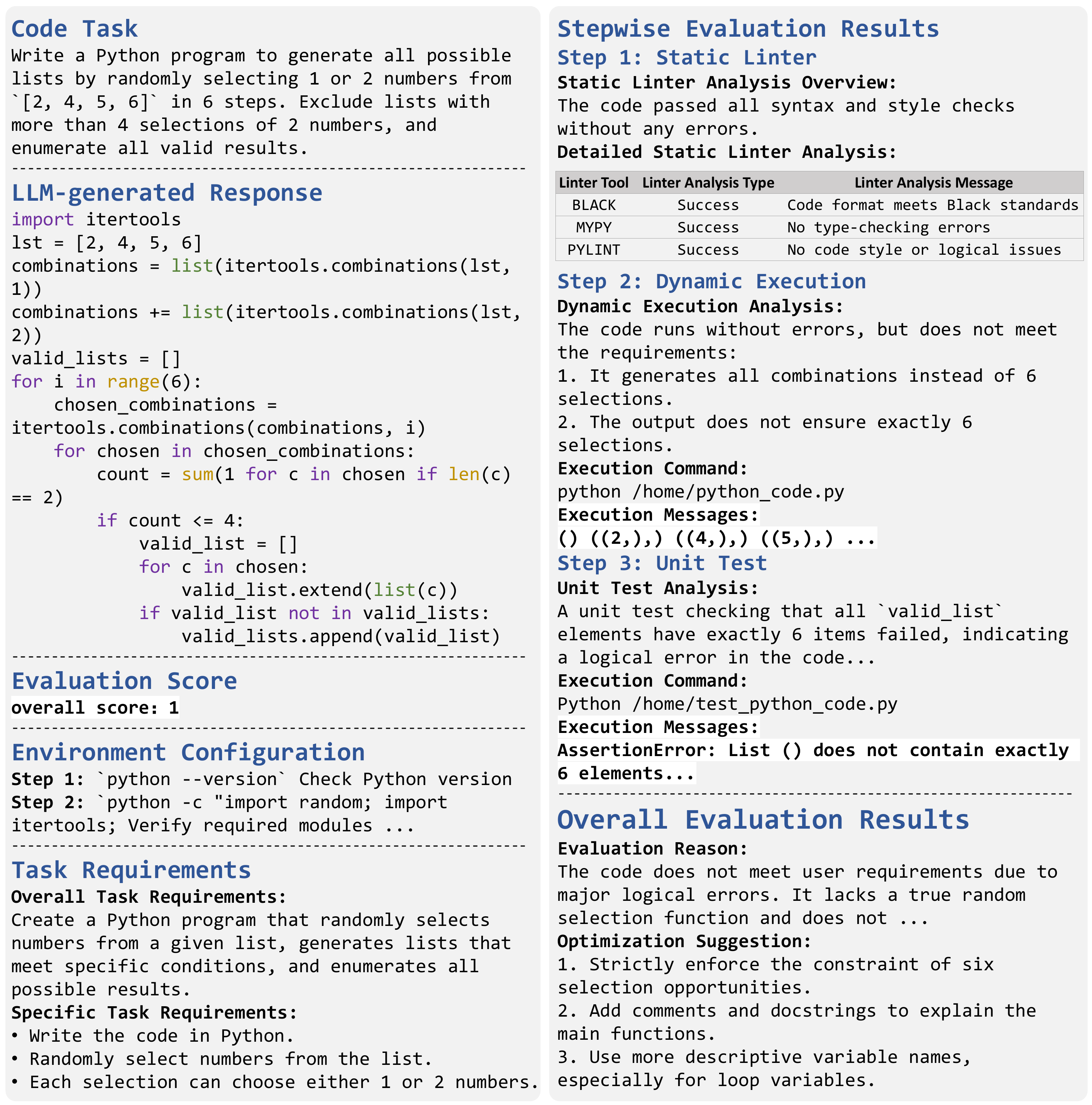}
    \vspace{-1em}
    \caption{Example evaluation report generated by \framework.}
\label{fig:report}
\end{figure}

As shown in Figures~\ref{fig:report}, the evaluation report begins with an overview of the ``Code Task'', the ``LLM-generated Response'', and the corresponding ``Evaluation Score''. The ``Environment Configuration'' section outlines the environment setup, facilitating reproducible and isolated evaluation. The ``Task Requirements'' section specifies both overall and specific requirements, enabling targeted assessment of functionality and completeness. The ``Stepwise Evaluation Results'' section provides a detailed, multi-stage assessment of the generated code. In this example, \framework utilizes ``Dynamic Execution'' and ``Unit Tests'' to identify discrepancies between the code execution output and the task requirements, specifically revealing that the length of the generated list does not match the expected length, thereby uncovering functional errors in the code. Besides, \framework utilizes ``Static Linter'' to check the code style for potential issues.  The ``Overall Evaluation Results'' section summarizes that the code fails to meet requirements due to major logical errors and offers clear suggestions, such as enforcing the six-selection limit and using more descriptive variable names.

\subsection{Performance on less complex benchmarks}

\begin{table}[t]
  \centering
    \setlength{\tabcolsep}{5.0mm}
  \caption{Comparison of performance on CoNaLa.}
  \begin{tabular}{lccc}
    \toprule
    \rowcolor{gray!25} Metrics & $r_p$ & $r_s$ & $\tau$ \\
    \midrule
    ICE-SCORE     & \textbf{0.655} & 0.596 & 0.534 \\
    CODEJUDGE    & 0.544 & 0.491 & 0.444 \\ \midrule
    \framework   & 0.644 & \textbf{0.637} & \textbf{0.572} \\ 
    \bottomrule
  \end{tabular}
  \label{tabs:conala}
\end{table}

To evaluate the performance of \framework on less complex tasks, we conduct experiments on the popular CoNaLa dataset~\cite{conala}. CoNaLa is a Python code generation benchmark consisting of 472 tasks collected from StackOverflow and 2,360 code snippets. We adopt the human annotation collected by Evtikhiev~\cite{human0} as ground truth. As shown in Table~\ref{tabs:conala}, \framework also achieves great performance on CoNaLa. Specifically, \framework surpasses the best baseline method by 0.041 and 0.038 on $r_s$ and $\tau$, respectively. Overall, the results indicate that \framework consistently maintains effectiveness across tasks with varying difficulty levels, showcasing its adaptability and scalability.

\subsection{Threats and Limitations}

One threat to validity is that our benchmark may not cover all coding scenarios and programming languages. Hence, the experimental results may not be fully generalizable. In the future, we intend to extend our benchmark to include more. Another threat arises from the inherent randomness of LLMs. Since \framework relies on LLM agents to evaluate and score responses, the results may vary across trials. We therefore perform multiple trials and take the average as the experimental results.

\section{Related Work}
\label{sec:related}
\subsection{Code Generation Evaluation}

Current approaches for code generation evaluation can be categorized into metric-based, human-centered, and LLM-based.
Early works directly transfer evaluation metrics from the natural language processing (NLP) domain to code generation, such as BLEU~\cite{bleu}, ROUGE~\cite{rogue}, and METEOR~\cite{meteor}. However, these metrics fail to capture the syntactic and functional correctness. To mitigate this limitation, CodeBLEU~\cite{codebleu} incorporates syntactic and semantic information from Abstract Syntax Trees (AST) and Data Flow Graphs (DFG). Recently, execution-based metrics have gained prominence, including pass@k~\cite{passk} and pass@t~\cite{passt}. However, these metrics are based on high-quality unit tests and are limited to executable code.

For more complex and diverse tasks~\cite{human1, human2}, human-centered approaches become particularly important. Human evaluators can provide reliable and accurate evaluations based on their professional knowledge and programming experience, which are also essential for fine-tuning LLMs to better align with human preferences~\cite{human3}. However, human-centered approaches are labour-intensive. Recently, LLM-based approaches~\cite{wang2025can} leverage LLMs to simulate the human evaluation process, enhancing efficiency and scalability. ICE-Score~\cite{ice-score} aligns well with human preferences in terms of code accuracy and functionality without test oracles or references. CODEJUDGE~\cite{codejudge} employs a ``slow thinking'' approach, enabling a thorough and reliable assessment of code's semantic correctness. 
Nevertheless, these LLM-based approaches still face limitations, including a lack of multi-dimensional context and omission of certain issues in complex code.

\subsection{LLM-Based Agents}

AI agents are artificial entities capable of accomplishing complex tasks by perceiving the environment and taking actions~\cite{DBLP:journals/corr/abs-2309-07864, DBLP:journals/corr/abs-2409-02977}. Recently, rapid progress in LLMs has greatly increased researchers' attention to LLM-based agents~\cite{DBLP:conf/iclr/ChenSZ0YCYLHQQC24,DBLP:conf/acl/Qiao0FLZJLC24}. LLM-based agent frameworks leverage LLM as the central controller to address complex tasks by integrating external tools and powerful understanding capabilities of LLMs. These agents typically consist of four core components: planning, memory, perception, and action~\cite{DBLP:journals/corr/abs-2309-07864}. The planning component ensures efficient task execution through strategies such as single or multi-planner approaches~\cite{plan3, plan4} and single or multi-turn planning~\cite{plan5, plan6}. The memory component retains historical data to support reasoning, with implementations ranging from short-term to long-term memory~\cite{memory1, DBLP:journals/corr/abs-2311-08649}, as well as specific or shared memory~\cite{memory3, memory4}. The perception component enables agents to process textual inputs~\cite{perception1, DBLP:journals/corr/abs-2311-08649} and visual data~\cite{perception3, perception4}. The action component integrates external tools~\cite{action1, action2} for tasks such as web searches, file operations, and GUI interactions, thereby extending the agents' functional capabilities.

LLM-based agent frameworks designed for software engineering have demonstrated superior performance across various software development and maintenance tasks~\cite{liu2024marscode, gao2025trae, hu2025llm, 10.1145/3696630.3728557, wen2024evalsva}, including requirements engineering~\cite{DBLP:journals/corr/abs-2405-03256, DBLP:journals/corr/abs-2310-13976}, code generation~\cite{DBLP:journals/corr/abs-2312-13010, DBLP:journals/corr/abs-2404-02183}, static bug detection~\cite{DBLP:journals/corr/abs-2403-14274, DBLP:journals/corr/abs-2310-08837}, code review~\cite{DBLP:journals/corr/abs-2402-02172, DBLP:journals/corr/abs-2404-18496}, unit testing~\cite{DBLP:journals/corr/abs-2305-04207}, system testing~\cite{DBLP:journals/corr/abs-2311-08649, DBLP:journals/corr/abs-2308-06782}, fault localization~\cite{DBLP:journals/corr/abs-2403-16362, DBLP:journals/corr/abs-2310-16340}, program repair~\cite{DBLP:journals/corr/abs-2403-17134}, end-to-end software development~\cite{DBLP:journals/corr/abs-2405-15793, autocoderover} and end-to-end software maintenance~\cite{DBLP:journals/corr/abs-2406-01304}. 


\section{Conclusion}
\label{sec:conclusion}
This paper focuses on evaluating code generation for complex code scenarios
and proposes a novel agent-based evaluation framework, named \framework. \framework consists of a \modulea for gathering multi-dimensional contextual information based on the decomposed task requirements and stepwise evaluation plan, and a \moduleb for employing multiple judges engaging in discussions to better comprehend complex code in a fine-grained manner, finally reaching a consensus on the evaluation score. Besides, we provide detailed evaluation reports assisting developers in identifying shortcomings. Compared with the state-of-the-art approaches, the experimental results validate the effectiveness of~\framework. In the future, we intend to further evaluate \framework on a broader range of datasets.


\section*{Acknowledgment}
This research is supported by the National Natural Science Foundation of China under project (No. 62472126, 62276075), Natural Science Foundation of Guangdong Province (Project No. 2023A1515011959), and Shenzhen-Hong Kong Jointly Funded Project (Category A, No. SGDX20230116 091246007).

\bibliographystyle{IEEEtran} 

\bibliography{IEEEabrv, Citation} 

\begin{thebibliography}{10}
\providecommand{\url}[1]{#1}
\csname url@samestyle\endcsname
\providecommand{\newblock}{\relax}
\providecommand{\bibinfo}[2]{#2}
\providecommand{\BIBentrySTDinterwordspacing}{\spaceskip=0pt\relax}
\providecommand{\BIBentryALTinterwordstretchfactor}{4}
\providecommand{\BIBentryALTinterwordspacing}{\spaceskip=\fontdimen2\font plus
\BIBentryALTinterwordstretchfactor\fontdimen3\font minus \fontdimen4\font\relax}
\providecommand{\BIBforeignlanguage}[2]{{%
\expandafter\ifx\csname l@#1\endcsname\relax
\typeout{** WARNING: IEEEtran.bst: No hyphenation pattern has been}%
\typeout{** loaded for the language `#1'. Using the pattern for}%
\typeout{** the default language instead.}%
\else
\language=\csname l@#1\endcsname
\fi
#2}}
\providecommand{\BIBdecl}{\relax}
\BIBdecl

\bibitem{codetask1}
B.~Roziere, J.~Gehring, F.~Gloeckle, S.~Sootla, I.~Gat, X.~E. Tan, Y.~Adi, J.~Liu, R.~Sauvestre, T.~Remez \emph{et~al.}, ``Code llama: Open foundation models for code,'' \emph{arXiv preprint arXiv:2308.12950}, 2023.

\bibitem{codetask2}
\BIBentryALTinterwordspacing
B.~Shen, J.~Zhang, T.~Chen, D.~Zan, B.~Geng, A.~Fu, M.~Zeng, A.~Yu, J.~Ji, J.~Zhao, Y.~Guo, and Q.~Wang, ``Pangu-coder2: Boosting large language models for code with ranking feedback,'' \emph{CoRR}, vol. abs/2307.14936, 2023. [Online]. Available: \url{https://doi.org/10.48550/arXiv.2307.14936}
\BIBentrySTDinterwordspacing

\bibitem{copilot}
GitHub, ``Github copilot - your ai pair programmer,'' \url{https://github. com/features/copilot}.

\bibitem{cursor}
A.~Cursor, ``{Cursor},'' \url{https://www.cursor.com/}.

\bibitem{human0}
\BIBentryALTinterwordspacing
M.~Evtikhiev, E.~Bogomolov, Y.~Sokolov, and T.~Bryksin, ``Out of the {BLEU:} how should we assess quality of the code generation models?'' \emph{J. Syst. Softw.}, vol. 203, p. 111741, 2023. [Online]. Available: \url{https://doi.org/10.1016/j.jss.2023.111741}
\BIBentrySTDinterwordspacing

\bibitem{llmsurvey2}
L.~Zheng, W.-L. Chiang, Y.~Sheng, S.~Zhuang, Z.~Wu, Y.~Zhuang, Z.~Lin, Z.~Li, D.~Li, E.~P. Xing, H.~Zhang, J.~E. Gonzalez, and I.~Stoica, ``Judging llm-as-a-judge with mt-bench and chatbot arena,'' in \emph{Proceedings of the 37th International Conference on Neural Information Processing Systems}, ser. NIPS '23.\hskip 1em plus 0.5em minus 0.4em\relax Red Hook, NY, USA: Curran Associates Inc., 2023.

\bibitem{bleu}
\BIBentryALTinterwordspacing
K.~Papineni, S.~Roukos, T.~Ward, and W.~Zhu, ``Bleu: a method for automatic evaluation of machine translation,'' in \emph{Proceedings of the 40th Annual Meeting of the Association for Computational Linguistics, July 6-12, 2002, Philadelphia, PA, {USA}}.\hskip 1em plus 0.5em minus 0.4em\relax {ACL}, 2002, pp. 311--318. [Online]. Available: \url{https://aclanthology.org/P02-1040/}
\BIBentrySTDinterwordspacing

\bibitem{codebleu}
\BIBentryALTinterwordspacing
S.~Ren, D.~Guo, S.~Lu, L.~Zhou, S.~Liu, D.~Tang, N.~Sundaresan, M.~Zhou, A.~Blanco, and S.~Ma, ``Codebleu: a method for automatic evaluation of code synthesis,'' \emph{CoRR}, vol. abs/2009.10297, 2020. [Online]. Available: \url{https://arxiv.org/abs/2009.10297}
\BIBentrySTDinterwordspacing

\bibitem{passk}
M.~Chen, J.~Tworek, H.~Jun, Q.~Yuan, H.~P. D.~O. Pinto, J.~Kaplan, H.~Edwards, Y.~Burda, N.~Joseph, G.~Brockman \emph{et~al.}, ``Evaluating large language models trained on code,'' \emph{arXiv preprint arXiv:2107.03374}, 2021.

\bibitem{passt}
T.~X. Olausson, J.~P. Inala, C.~Wang, J.~Gao, and A.~Solar-Lezama, ``Is self-repair a silver bullet for code generation?'' in \emph{International Conference on Learning Representations (ICLR)}, 2024.

\bibitem{exec-acc}
\BIBentryALTinterwordspacing
N.~Rajkumar, R.~Li, and D.~Bahdanau, ``Evaluating the text-to-sql capabilities of large language models,'' \emph{CoRR}, vol. abs/2204.00498, 2022. [Online]. Available: \url{https://doi.org/10.48550/arXiv.2204.00498}
\BIBentrySTDinterwordspacing

\bibitem{llmsurvey}
\BIBentryALTinterwordspacing
L.~Chen, Q.~Guo, H.~Jia, Z.~Zeng, X.~Wang, Y.~Xu, J.~Wu, Y.~Wang, Q.~Gao, J.~Wang, W.~Ye, and S.~Zhang, ``A survey on evaluating large language models in code generation tasks,'' \emph{CoRR}, vol. abs/2408.16498, 2024. [Online]. Available: \url{https://doi.org/10.48550/arXiv.2408.16498}
\BIBentrySTDinterwordspacing

\bibitem{llmsurvey3}
\BIBentryALTinterwordspacing
J.~Jiang, F.~Wang, J.~Shen, S.~Kim, and S.~Kim, ``A survey on large language models for code generation,'' \emph{CoRR}, vol. abs/2406.00515, 2024. [Online]. Available: \url{https://doi.org/10.48550/arXiv.2406.00515}
\BIBentrySTDinterwordspacing

\bibitem{ice-score}
\BIBentryALTinterwordspacing
T.~Y. Zhuo, ``Ice-score: Instructing large language models to evaluate code,'' in \emph{Findings of the Association for Computational Linguistics: {EACL} 2024, St. Julian's, Malta, March 17-22, 2024}, Y.~Graham and M.~Purver, Eds.\hskip 1em plus 0.5em minus 0.4em\relax Association for Computational Linguistics, 2024, pp. 2232--2242. [Online]. Available: \url{https://aclanthology.org/2024.findings-eacl.148}
\BIBentrySTDinterwordspacing

\bibitem{codejudge}
\BIBentryALTinterwordspacing
W.~Tong and T.~Zhang, ``Codejudge: Evaluating code generation with large language models,'' in \emph{Proceedings of the 2024 Conference on Empirical Methods in Natural Language Processing, {EMNLP} 2024, Miami, FL, USA, November 12-16, 2024}, Y.~Al{-}Onaizan, M.~Bansal, and Y.~Chen, Eds.\hskip 1em plus 0.5em minus 0.4em\relax Association for Computational Linguistics, 2024, pp. 20\,032--20\,051. [Online]. Available: \url{https://aclanthology.org/2024.emnlp-main.1118}
\BIBentrySTDinterwordspacing

\bibitem{llmsurvey1}
\BIBentryALTinterwordspacing
P.~Wang, L.~Li, L.~Chen, Z.~Cai, D.~Zhu, B.~Lin, Y.~Cao, L.~Kong, Q.~Liu, T.~Liu, and Z.~Sui, ``Large language models are not fair evaluators,'' in \emph{Proceedings of the 62nd Annual Meeting of the Association for Computational Linguistics (Volume 1: Long Papers), {ACL} 2024, Bangkok, Thailand, August 11-16, 2024}, L.~Ku, A.~Martins, and V.~Srikumar, Eds.\hskip 1em plus 0.5em minus 0.4em\relax Association for Computational Linguistics, 2024, pp. 9440--9450. [Online]. Available: \url{https://doi.org/10.18653/v1/2024.acl-long.511}
\BIBentrySTDinterwordspacing

\bibitem{DBLP:journals/corr/abs-2405-03256}
\BIBentryALTinterwordspacing
D.~Jin, Z.~Jin, X.~Chen, and C.~Wang, ``{MARE:} multi-agents collaboration framework for requirements engineering,'' \emph{CoRR}, vol. abs/2405.03256, 2024. [Online]. Available: \url{https://doi.org/10.48550/arXiv.2405.03256}
\BIBentrySTDinterwordspacing

\bibitem{DBLP:journals/corr/abs-2312-13010}
\BIBentryALTinterwordspacing
D.~Huang, Q.~Bu, J.~M. Zhang, M.~Luck, and H.~Cui, ``Agentcoder: Multi-agent-based code generation with iterative testing and optimisation,'' \emph{CoRR}, vol. abs/2312.13010, 2023. [Online]. Available: \url{https://doi.org/10.48550/arXiv.2312.13010}
\BIBentrySTDinterwordspacing

\bibitem{DBLP:journals/corr/abs-2403-14274}
Z.~Mao, J.~Li, D.~Jin, M.~Li, and K.~Tei, ``Multi-role consensus through llms discussions for vulnerability detection,'' in \emph{2024 IEEE 24th International Conference on Software Quality, Reliability, and Security Companion (QRS-C)}.\hskip 1em plus 0.5em minus 0.4em\relax IEEE, 2024, pp. 1318--1319.

\bibitem{plan1}
\BIBentryALTinterwordspacing
L.~Wang, W.~Xu, Y.~Lan, Z.~Hu, Y.~Lan, R.~K. Lee, and E.~Lim, ``Plan-and-solve prompting: Improving zero-shot chain-of-thought reasoning by large language models,'' in \emph{Proceedings of the 61st Annual Meeting of the Association for Computational Linguistics (Volume 1: Long Papers), {ACL} 2023, Toronto, Canada, July 9-14, 2023}, A.~Rogers, J.~L. Boyd{-}Graber, and N.~Okazaki, Eds.\hskip 1em plus 0.5em minus 0.4em\relax Association for Computational Linguistics, 2023, pp. 2609--2634. [Online]. Available: \url{https://doi.org/10.18653/v1/2023.acl-long.147}
\BIBentrySTDinterwordspacing

\bibitem{plan2}
\BIBentryALTinterwordspacing
X.~Jiang, Y.~Dong, L.~Wang, Z.~Fang, Q.~Shang, G.~Li, Z.~Jin, and W.~Jiao, ``Self-planning code generation with large language models,'' \emph{{ACM} Trans. Softw. Eng. Methodol.}, vol.~33, no.~7, pp. 182:1--182:30, 2024. [Online]. Available: \url{https://doi.org/10.1145/3672456}
\BIBentrySTDinterwordspacing

\bibitem{prettier}
prettier, ``{Prettier},'' \url{https://github.com/prettier/prettier}.

\bibitem{pandoc}
jgm, ``{Pandoc},'' \url{https://github.com/jgm/pandoc}.

\bibitem{codearena}
\BIBentryALTinterwordspacing
J.~Yang, J.~Yang, K.~Jin, Y.~Miao, L.~Zhang, L.~Yang, Z.~Cui, Y.~Zhang, B.~Hui, and J.~Lin, ``Evaluating and aligning codellms on human preference,'' \emph{CoRR}, vol. abs/2412.05210, 2024. [Online]. Available: \url{https://doi.org/10.48550/arXiv.2412.05210}
\BIBentrySTDinterwordspacing

\bibitem{gpt35}
OpenAI, ``{GPT-3.5 Turbo fine-tuning and API updates},'' \url{https://openai.com/index/gpt-3-5-turbo-fine-tuning-and-api-updates/}.

\bibitem{claude}
Anthropic, ``{Claude 3.5 Sonnet},'' \url{https://www.anthropic.com/news/claude-3-5-sonnet}.

\bibitem{gpt4O}
OpenAI, ``{Hello GPT-4o},'' \url{https://openai.com/index/hello-gpt-4o/}.

\bibitem{scipy}
scipy, ``{SciPy},'' \url{https://github.com/scipy/scipy}.

\bibitem{conala}
P.~Yin, B.~Deng, E.~Chen, B.~Vasilescu, and G.~Neubig, ``Learning to mine aligned code and natural language pairs from stack overflow,'' in \emph{Proceedings of the 15th international conference on mining software repositories}, 2018, pp. 476--486.

\bibitem{rogue}
\BIBentryALTinterwordspacing
C.-Y. Lin, ``Rouge: A package for automatic evaluation of summaries,'' in \emph{Annual Meeting of the Association for Computational Linguistics}, 2004. [Online]. Available: \url{https://api.semanticscholar.org/CorpusID:964287}
\BIBentrySTDinterwordspacing

\bibitem{meteor}
\BIBentryALTinterwordspacing
S.~Banerjee and A.~Lavie, ``{METEOR:} an automatic metric for {MT} evaluation with improved correlation with human judgments,'' in \emph{Proceedings of the Workshop on Intrinsic and Extrinsic Evaluation Measures for Machine Translation and/or Summarization@ACL 2005, Ann Arbor, Michigan, USA, June 29, 2005}, J.~Goldstein, A.~Lavie, C.~Lin, and C.~R. Voss, Eds.\hskip 1em plus 0.5em minus 0.4em\relax Association for Computational Linguistics, 2005, pp. 65--72. [Online]. Available: \url{https://aclanthology.org/W05-0909/}
\BIBentrySTDinterwordspacing

\bibitem{human1}
\BIBentryALTinterwordspacing
R.~Bairi, A.~Sonwane, A.~Kanade, V.~D. C., A.~Iyer, S.~Parthasarathy, S.~K. Rajamani, B.~Ashok, and S.~Shet, ``Codeplan: Repository-level coding using llms and planning,'' \emph{Proc. {ACM} Softw. Eng.}, vol.~1, no. {FSE}, pp. 675--698, 2024. [Online]. Available: \url{https://doi.org/10.1145/3643757}
\BIBentrySTDinterwordspacing

\bibitem{human2}
\BIBentryALTinterwordspacing
D.~Shrivastava, D.~Kocetkov, H.~de~Vries, D.~Bahdanau, and T.~Scholak, ``Repofusion: Training code models to understand your repository,'' \emph{CoRR}, vol. abs/2306.10998, 2023. [Online]. Available: \url{https://doi.org/10.48550/arXiv.2306.10998}
\BIBentrySTDinterwordspacing

\bibitem{human3}
L.~Ouyang, J.~Wu, X.~Jiang, D.~Almeida, C.~L. Wainwright, P.~Mishkin, C.~Zhang, S.~Agarwal, K.~Slama, A.~Ray, J.~Schulman, J.~Hilton, F.~Kelton, L.~Miller, M.~Simens, A.~Askell, P.~Welinder, P.~Christiano, J.~Leike, and R.~Lowe, ``Training language models to follow instructions with human feedback,'' in \emph{Proceedings of the 36th International Conference on Neural Information Processing Systems}, ser. NIPS '22.\hskip 1em plus 0.5em minus 0.4em\relax Red Hook, NY, USA: Curran Associates Inc., 2022.

\bibitem{wang2025can}
R.~Wang, J.~Guo, C.~Gao, G.~Fan, C.~Y. Chong, and X.~Xia, ``Can llms replace human evaluators? an empirical study of llm-as-a-judge in software engineering,'' \emph{Proceedings of the ACM on Software Engineering}, vol.~2, no. ISSTA, pp. 1955--1977, 2025.

\bibitem{DBLP:journals/corr/abs-2309-07864}
Z.~Xi, W.~Chen, X.~Guo, W.~He, Y.~Ding, B.~Hong, M.~Zhang, J.~Wang, S.~Jin, E.~Zhou \emph{et~al.}, ``The rise and potential of large language model based agents: A survey,'' \emph{Science China Information Sciences}, vol.~68, no.~2, p. 121101, 2025.

\bibitem{DBLP:journals/corr/abs-2409-02977}
\BIBentryALTinterwordspacing
J.~Liu, K.~Wang, Y.~Chen, X.~Peng, Z.~Chen, L.~Zhang, and Y.~Lou, ``Large language model-based agents for software engineering: {A} survey,'' \emph{CoRR}, vol. abs/2409.02977, 2024. [Online]. Available: \url{https://doi.org/10.48550/arXiv.2409.02977}
\BIBentrySTDinterwordspacing

\bibitem{DBLP:conf/iclr/ChenSZ0YCYLHQQC24}
\BIBentryALTinterwordspacing
W.~Chen, Y.~Su, J.~Zuo, C.~Yang, C.~Yuan, C.~Chan, H.~Yu, Y.~Lu, Y.~Hung, C.~Qian, Y.~Qin, X.~Cong, R.~Xie, Z.~Liu, M.~Sun, and J.~Zhou, ``Agentverse: Facilitating multi-agent collaboration and exploring emergent behaviors,'' in \emph{The Twelfth International Conference on Learning Representations, {ICLR} 2024, Vienna, Austria, May 7-11, 2024}.\hskip 1em plus 0.5em minus 0.4em\relax OpenReview.net, 2024. [Online]. Available: \url{https://openreview.net/forum?id=EHg5GDnyq1}
\BIBentrySTDinterwordspacing

\bibitem{DBLP:conf/acl/Qiao0FLZJLC24}
\BIBentryALTinterwordspacing
S.~Qiao, N.~Zhang, R.~Fang, Y.~Luo, W.~Zhou, Y.~E. Jiang, C.~Lv, and H.~Chen, ``Autoact: Automatic agent learning from scratch for {QA} via self-planning,'' in \emph{Proceedings of the 62nd Annual Meeting of the Association for Computational Linguistics (Volume 1: Long Papers), {ACL} 2024, Bangkok, Thailand, August 11-16, 2024}, L.~Ku, A.~Martins, and V.~Srikumar, Eds.\hskip 1em plus 0.5em minus 0.4em\relax Association for Computational Linguistics, 2024, pp. 3003--3021. [Online]. Available: \url{https://aclanthology.org/2024.acl-long.165}
\BIBentrySTDinterwordspacing

\bibitem{plan3}
\BIBentryALTinterwordspacing
Y.~Dong, X.~Jiang, Z.~Jin, and G.~Li, ``Self-collaboration code generation via chatgpt,'' \emph{{ACM} Trans. Softw. Eng. Methodol.}, vol.~33, no.~7, pp. 189:1--189:38, 2024. [Online]. Available: \url{https://doi.org/10.1145/3672459}
\BIBentrySTDinterwordspacing

\bibitem{plan4}
\BIBentryALTinterwordspacing
M.~Josifoski, L.~H. Klein, M.~Peyrard, Y.~Li, S.~Geng, J.~P. Schnitzler, Y.~Yao, J.~Wei, D.~Paul, and R.~West, ``Flows: Building blocks of reasoning and collaborating {AI},'' \emph{CoRR}, vol. abs/2308.01285, 2023. [Online]. Available: \url{https://doi.org/10.48550/arXiv.2308.01285}
\BIBentrySTDinterwordspacing

\bibitem{plan5}
\BIBentryALTinterwordspacing
E.~Zelikman, Q.~Huang, G.~Poesia, N.~D. Goodman, and N.~Haber, ``Parsel: Algorithmic reasoning with language models by composing decompositions,'' in \emph{Advances in Neural Information Processing Systems 36: Annual Conference on Neural Information Processing Systems 2023, NeurIPS 2023, New Orleans, LA, USA, December 10 - 16, 2023}, A.~Oh, T.~Naumann, A.~Globerson, K.~Saenko, M.~Hardt, and S.~Levine, Eds., 2023. [Online]. Available: \url{https://dl.acm.org/doi/10.5555/3666122.3667489}
\BIBentrySTDinterwordspacing

\bibitem{plan6}
A.~Zhou, K.~Yan, M.~Shlapentokh-Rothman, H.~Wang, and Y.-X. Wang, ``Language agent tree search unifies reasoning, acting, and planning in language models,'' in \emph{Proceedings of the 41st International Conference on Machine Learning}, ser. ICML'24.\hskip 1em plus 0.5em minus 0.4em\relax JMLR.org, 2024.

\bibitem{memory1}
\BIBentryALTinterwordspacing
M.~Geva, R.~Schuster, J.~Berant, and O.~Levy, ``Transformer feed-forward layers are key-value memories,'' in \emph{Proceedings of the 2021 Conference on Empirical Methods in Natural Language Processing, {EMNLP} 2021, Virtual Event / Punta Cana, Dominican Republic, 7-11 November, 2021}, M.~Moens, X.~Huang, L.~Specia, and S.~W. Yih, Eds.\hskip 1em plus 0.5em minus 0.4em\relax Association for Computational Linguistics, 2021, pp. 5484--5495. [Online]. Available: \url{https://doi.org/10.18653/v1/2021.emnlp-main.446}
\BIBentrySTDinterwordspacing

\bibitem{DBLP:journals/corr/abs-2311-08649}
\BIBentryALTinterwordspacing
J.~Yoon, R.~Feldt, and S.~Yoo, ``Autonomous large language model agents enabling intent-driven mobile {GUI} testing,'' \emph{CoRR}, vol. abs/2311.08649, 2023. [Online]. Available: \url{https://doi.org/10.48550/arXiv.2311.08649}
\BIBentrySTDinterwordspacing

\bibitem{memory3}
\BIBentryALTinterwordspacing
C.~Qian, W.~Liu, H.~Liu, N.~Chen, Y.~Dang, J.~Li, C.~Yang, W.~Chen, Y.~Su, X.~Cong, J.~Xu, D.~Li, Z.~Liu, and M.~Sun, ``Chatdev: Communicative agents for software development,'' in \emph{Proceedings of the 62nd Annual Meeting of the Association for Computational Linguistics (Volume 1: Long Papers), {ACL} 2024, Bangkok, Thailand, August 11-16, 2024}, L.~Ku, A.~Martins, and V.~Srikumar, Eds.\hskip 1em plus 0.5em minus 0.4em\relax Association for Computational Linguistics, 2024, pp. 15\,174--15\,186. [Online]. Available: \url{https://doi.org/10.18653/v1/2024.acl-long.810}
\BIBentrySTDinterwordspacing

\bibitem{memory4}
\BIBentryALTinterwordspacing
S.~Hong, M.~Zhuge, J.~Chen, X.~Zheng, Y.~Cheng, J.~Wang, C.~Zhang, Z.~Wang, S.~K.~S. Yau, Z.~Lin, L.~Zhou, C.~Ran, L.~Xiao, C.~Wu, and J.~Schmidhuber, ``Metagpt: Meta programming for {A} multi-agent collaborative framework,'' in \emph{The Twelfth International Conference on Learning Representations, {ICLR} 2024, Vienna, Austria, May 7-11, 2024}.\hskip 1em plus 0.5em minus 0.4em\relax OpenReview.net, 2024. [Online]. Available: \url{https://openreview.net/forum?id=VtmBAGCN7o}
\BIBentrySTDinterwordspacing

\bibitem{perception1}
\BIBentryALTinterwordspacing
J.~A. Pizzorno and E.~D. Berger, ``Coverup: Coverage-guided llm-based test generation,'' \emph{CoRR}, vol. abs/2403.16218, 2024. [Online]. Available: \url{https://doi.org/10.48550/arXiv.2403.16218}
\BIBentrySTDinterwordspacing

\bibitem{perception3}
\BIBentryALTinterwordspacing
Z.~Wang, W.~Wang, Z.~Li, L.~Wang, C.~Yi, X.~Xu, L.~Cao, H.~Su, S.~Chen, and J.~Zhou, ``Xuat-copilot: Multi-agent collaborative system for automated user acceptance testing with large language model,'' \emph{CoRR}, vol. abs/2401.02705, 2024. [Online]. Available: \url{https://doi.org/10.48550/arXiv.2401.02705}
\BIBentrySTDinterwordspacing

\bibitem{perception4}
\BIBentryALTinterwordspacing
M.~Taeb, A.~Swearngin, E.~Schoop, R.~Cheng, Y.~Jiang, and J.~Nichols, ``Axnav: Replaying accessibility tests from natural language,'' in \emph{Proceedings of the {CHI} Conference on Human Factors in Computing Systems, {CHI} 2024, Honolulu, HI, USA, May 11-16, 2024}, F.~F. Mueller, P.~Kyburz, J.~R. Williamson, C.~Sas, M.~L. Wilson, P.~O.~T. Dugas, and I.~Shklovski, Eds.\hskip 1em plus 0.5em minus 0.4em\relax {ACM}, 2024, pp. 962:1--962:16. [Online]. Available: \url{https://doi.org/10.1145/3613904.3642777}
\BIBentrySTDinterwordspacing

\bibitem{action1}
\BIBentryALTinterwordspacing
K.~Zhang, G.~Li, J.~Li, Z.~Li, and Z.~Jin, ``Toolcoder: Teach code generation models to use {API} search tools,'' \emph{CoRR}, vol. abs/2305.04032, 2023. [Online]. Available: \url{https://doi.org/10.48550/arXiv.2305.04032}
\BIBentrySTDinterwordspacing

\bibitem{action2}
\BIBentryALTinterwordspacing
Z.~Li, S.~Dutta, and M.~Naik, ``Llm-assisted static analysis for detecting security vulnerabilities,'' \emph{CoRR}, vol. abs/2405.17238, 2024. [Online]. Available: \url{https://doi.org/10.48550/arXiv.2405.17238}
\BIBentrySTDinterwordspacing

\bibitem{liu2024marscode}
Y.~Liu, P.~Gao, X.~Wang, J.~Liu, Y.~Shi, Z.~Zhang, and C.~Peng, ``Marscode agent: Ai-native automated bug fixing,'' \emph{arXiv preprint arXiv:2409.00899}, 2024.

\bibitem{gao2025trae}
P.~Gao, Z.~Tian, X.~Meng, X.~Wang, R.~Hu, Y.~Xiao, Y.~Liu, Z.~Zhang, J.~Chen, C.~Gao \emph{et~al.}, ``Trae agent: An llm-based agent for software engineering with test-time scaling,'' \emph{arXiv preprint arXiv:2507.23370}, 2025.

\bibitem{hu2025llm}
R.~Hu, C.~Peng, X.~Wang, and C.~Gao, ``An llm-based agent for reliable docker environment configuration,'' \emph{arXiv preprint arXiv:2502.13681}, 2025.

\bibitem{10.1145/3696630.3728557}
\BIBentryALTinterwordspacing
X.~Wang, P.~Gao, X.~Meng, C.~Peng, R.~Hu, Y.~Lin, and C.~Gao, ``Aegis: An agent-based framework for bug reproduction from issue descriptions,'' in \emph{Proceedings of the 33rd ACM International Conference on the Foundations of Software Engineering}, ser. FSE Companion '25.\hskip 1em plus 0.5em minus 0.4em\relax New York, NY, USA: Association for Computing Machinery, 2025, p. 331–342. [Online]. Available: \url{https://doi.org/10.1145/3696630.3728557}
\BIBentrySTDinterwordspacing

\bibitem{wen2024evalsva}
X.-C. Wen, J.~Ye, C.~Gao, L.~Wu, and Q.~Liao, ``Evalsva: Multi-agent evaluators for next-gen software vulnerability assessment,'' \emph{arXiv preprint arXiv:2501.14737}, 2024.

\bibitem{DBLP:journals/corr/abs-2310-13976}
C.~Arora, J.~Grundy, and M.~Abdelrazek, ``Advancing requirements engineering through generative ai: Assessing the role of llms,'' in \emph{Generative AI for Effective Software Development}.\hskip 1em plus 0.5em minus 0.4em\relax Springer, 2024, pp. 129--148.

\bibitem{DBLP:journals/corr/abs-2404-02183}
\BIBentryALTinterwordspacing
Y.~Ishibashi and Y.~Nishimura, ``Self-organized agents: {A} {LLM} multi-agent framework toward ultra large-scale code generation and optimization,'' \emph{CoRR}, vol. abs/2404.02183, 2024. [Online]. Available: \url{https://doi.org/10.48550/arXiv.2404.02183}
\BIBentrySTDinterwordspacing

\bibitem{DBLP:journals/corr/abs-2310-08837}
\BIBentryALTinterwordspacing
G.~Fan, X.~Xie, X.~Zheng, Y.~Liang, and P.~Di, ``Static code analysis in the {AI} era: An in-depth exploration of the concept, function, and potential of intelligent code analysis agents,'' \emph{CoRR}, vol. abs/2310.08837, 2023. [Online]. Available: \url{https://doi.org/10.48550/arXiv.2310.08837}
\BIBentrySTDinterwordspacing

\bibitem{DBLP:journals/corr/abs-2402-02172}
\BIBentryALTinterwordspacing
D.~Tang, Z.~Chen, K.~Kim, Y.~Song, H.~Tian, S.~Ezzini, Y.~Huang, J.~Klein, and T.~F. Bissyand{\'{e}}, ``Codeagent: Collaborative agents for software engineering,'' \emph{CoRR}, vol. abs/2402.02172, 2024. [Online]. Available: \url{https://doi.org/10.48550/arXiv.2402.02172}
\BIBentrySTDinterwordspacing

\bibitem{DBLP:journals/corr/abs-2404-18496}
\BIBentryALTinterwordspacing
Z.~Rasheed, M.~A. Sami, M.~Waseem, K.~Kemell, X.~Wang, A.~Nguyen, K.~Syst{\"{a}}, and P.~Abrahamsson, ``Ai-powered code review with llms: Early results,'' \emph{CoRR}, vol. abs/2404.18496, 2024. [Online]. Available: \url{https://doi.org/10.48550/arXiv.2404.18496}
\BIBentrySTDinterwordspacing

\bibitem{DBLP:journals/corr/abs-2305-04207}
\BIBentryALTinterwordspacing
Z.~Yuan, Y.~Lou, M.~Liu, S.~Ding, K.~Wang, Y.~Chen, and X.~Peng, ``No more manual tests? evaluating and improving chatgpt for unit test generation,'' \emph{CoRR}, vol. abs/2305.04207, 2023. [Online]. Available: \url{https://doi.org/10.48550/arXiv.2305.04207}
\BIBentrySTDinterwordspacing

\bibitem{DBLP:journals/corr/abs-2308-06782}
\BIBentryALTinterwordspacing
G.~Deng, Y.~Liu, V.~M. Vilches, P.~Liu, Y.~Li, Y.~Xu, M.~Pinzger, S.~Rass, T.~Zhang, and Y.~Liu, ``Pentestgpt: Evaluating and harnessing large language models for automated penetration testing,'' in \emph{33rd {USENIX} Security Symposium, {USENIX} Security 2024, Philadelphia, PA, USA, August 14-16, 2024}, D.~Balzarotti and W.~Xu, Eds.\hskip 1em plus 0.5em minus 0.4em\relax {USENIX} Association, 2024. [Online]. Available: \url{https://www.usenix.org/conference/usenixsecurity24/presentation/deng}
\BIBentrySTDinterwordspacing

\bibitem{DBLP:journals/corr/abs-2403-16362}
\BIBentryALTinterwordspacing
Y.~Qin, S.~Wang, Y.~Lou, J.~Dong, K.~Wang, X.~Li, and X.~Mao, ``Agentfl: Scaling llm-based fault localization to project-level context,'' \emph{CoRR}, vol. abs/2403.16362, 2024. [Online]. Available: \url{https://doi.org/10.48550/arXiv.2403.16362}
\BIBentrySTDinterwordspacing

\bibitem{DBLP:journals/corr/abs-2310-16340}
Z.~Wang, Z.~Liu, Y.~Zhang, A.~Zhong, J.~Wang, F.~Yin, L.~Fan, L.~Wu, and Q.~Wen, ``Rcagent: Cloud root cause analysis by autonomous agents with tool-augmented large language models,'' in \emph{Proceedings of the 33rd ACM International Conference on Information and Knowledge Management}, 2024, pp. 4966--4974.

\bibitem{DBLP:journals/corr/abs-2403-17134}
\BIBentryALTinterwordspacing
I.~Bouzenia, P.~T. Devanbu, and M.~Pradel, ``Repairagent: An autonomous, llm-based agent for program repair,'' \emph{CoRR}, vol. abs/2403.17134, 2024. [Online]. Available: \url{https://doi.org/10.48550/arXiv.2403.17134}
\BIBentrySTDinterwordspacing

\bibitem{DBLP:journals/corr/abs-2405-15793}
J.~Yang, C.~E. Jimenez, A.~Wettig, K.~Lieret, S.~Yao, K.~Narasimhan, and O.~Press, ``Swe-agent: Agent-computer interfaces enable automated software engineering,'' \emph{Advances in Neural Information Processing Systems}, vol.~37, pp. 50\,528--50\,652, 2024.

\bibitem{autocoderover}
\BIBentryALTinterwordspacing
Y.~Zhang, H.~Ruan, Z.~Fan, and A.~Roychoudhury, ``Autocoderover: Autonomous program improvement,'' in \emph{Proceedings of the 33rd {ACM} {SIGSOFT} International Symposium on Software Testing and Analysis, {ISSTA} 2024, Vienna, Austria, September 16-20, 2024}, M.~Christakis and M.~Pradel, Eds.\hskip 1em plus 0.5em minus 0.4em\relax {ACM}, 2024, pp. 1592--1604. [Online]. Available: \url{https://doi.org/10.1145/3650212.3680384}
\BIBentrySTDinterwordspacing

\bibitem{DBLP:journals/corr/abs-2406-01304}
\BIBentryALTinterwordspacing
D.~Chen, S.~Lin, M.~Zeng, D.~Zan, J.~Wang, A.~Cheshkov, J.~Sun, H.~Yu, G.~Dong, A.~Aliev, J.~Wang, X.~Cheng, G.~Liang, Y.~Ma, P.~Bian, T.~Xie, and Q.~Wang, ``Coder: Issue resolving with multi-agent and task graphs,'' \emph{CoRR}, vol. abs/2406.01304, 2024. [Online]. Available: \url{https://doi.org/10.48550/arXiv.2406.01304}
\BIBentrySTDinterwordspacing

\end{thebibliography}

\end{document}